\documentclass[letterpaper, 10 pt, conference]{ieeeconf}

\IEEEoverridecommandlockouts                              

\usepackage{empheq}
\usepackage{amsmath,amssymb,amsfonts}
\usepackage{environ}
\usepackage{cite}
\usepackage{grffile}
\usepackage{multicol}
\usepackage{graphics}
\usepackage{subcaption}
\usepackage{url}
\usepackage{float}

\title{\LARGE \bf
Comparing Spatially Periodic Feedback and Space-Time Modulation for Unidirectional Wave Propagation in a 1D Mass-Spring-Damper System} 
\author{João Henrique Sousa Brandão$^{1}$, Danilo Braghini$^{2}$, José Roberto de França Arruda$^{1}$
\thanks{$^{1}$Danilo Braghini\{{\tt\small dbraghin@asu.edu}\} is with the School for Engineering of Matter, Transport and Energy at Arizona State University, Tempe, AZ, USA.}
	\thanks{João Henrique Sousa Brandão \{{\tt\small j174253@dac.unicamp.br}\}, and José Roberto de França Arruda\{ {arruda@fem.unicamp.br}\} are with the School of Mechanical Engineering, University of Campinas, Campinas, São Paulo 13083-970, Brazil
	}
    }

\begin{document}
	\maketitle
	\thispagestyle{empty}
	\pagestyle{empty}
	
\section{INTRODUCTION}

Wave propagation in periodic structures has been extensively studied 
to manipulate waves through phenomena such as band gaps, whereby wave propagation is prohibited within specific frequency ranges, 
and 
non-reciprocity, 
whereby interchanging the excitation and the observation points results in different dynamic responses. Although linear, passive, and time-invariant physical systems present reciprocal dynamic responses, reciprocity can be broken, e.g., by time-varying properties or external action, giving rise to 
directional band gaps and wave modes with unidirectional amplification or attenuation~\cite{Coulais2017, Nassar2017, Trainiti2016}. Non-reciprocal systems have 
potential applications in 
the development of acoustic and mechanical diodes~\cite{Wang2018}, non-reciprocal circulators~\cite{Fleury2014}, 
directional vibration insulation and energy harvesting~\cite{Wang2018, Coulais2017}, 
parametric amplification and frequency conversion~\cite{Ye2025}.

Several design approaches have been proposed to achieve non-reciprocity in mechanical systems: \cite{BELI20181077, Fleury2014} showed that non-reciprocal behavior can be obtained by introducing circulating flows or rotating bodies in acoustic and mechanical systems, respectively; \cite{Lepri2011} proposed a system that combines nonlinear material behavior with geometric asymmetry to realize non-reciprocity; \cite{Nassar2017, VILA2017363} showed that non-reciprocity can be achieved through space-time modulated (STM) system properties; and~\cite{Braghini_2021, Rosa_2020} presented systems with spatially periodic active feedback interactions that leverages topologically protected non-reciprocal dynamics. While 
most of the approaches generate non-reciprocity by breaking time-reversal symmetry, the feedback approach relies on the Non-Hermitian Skin Effect (NHSE) to achieve non-reciprocal wave propagation. 
This work focuses on comparing the STM and feedback approaches, which can be implemented in simple 
one-dimensional mass-spring-damper linear system. Then, the dynamic analysis is straightforward. 
Specifically, our results are presented twofold: 
local in space and transient in time, through the analysis of wave propagation; global in space and permanent in time, through stability analysis.

In this work, we compare non-reciprocal wave propagation in a one-dimensional, linear, spatially periodic lumped-parameter mechanical system (mass-spring-damper) within the Floquet-Bloch framework for two distinct classes of non-reciprocal systems: (i)~a spatially periodic system with constant properties and a concentrated feedback control law, and (ii)~a system with space-time modulated properties. The objective is to analyze and compare the non-reciprocal effects in both systems, highlighting similarities and differences in their wave propagation characteristics, and to discuss the advantages and limitations of each approach for practical applications.

Both systems are modeled using the Plane Wave Expansion (PWE) formulation within Floquet-Bloch theory, a widely used technique for analyzing wave propagation in periodic systems~\cite{Vasseur2021}. The dispersion diagram, a graphical representation of the frequency-wavenumber relationship, serves as the primary tool for comparing the non-reciprocal effects, 
enabling the identification of band gaps, passbands, and frequency ranges that exhibit unidirectional attenuation or amplification. The corresponding finite structures are used to simulate the dynamic response to a localized transient excitation with a strategically chosen frequency that highlights the effects 
presented by the dispersion diagrams. The finite structure dynamic response 
in the presence of directional band gaps or unidirectional amplification/attenuation 
serve as clear indicators of non-reciprocity. Additionally, stability is assessed for both configurations in the search for parameters -- modulation amplitude/frequency and feedback gain -- that yield stable structures. 

\section{THEORETICAL BACKGROUND}

\subsection{Feedback-induced non-reciprocity}

\begin{figure*}[h!]
\centering
\includegraphics[width=0.9\textwidth]{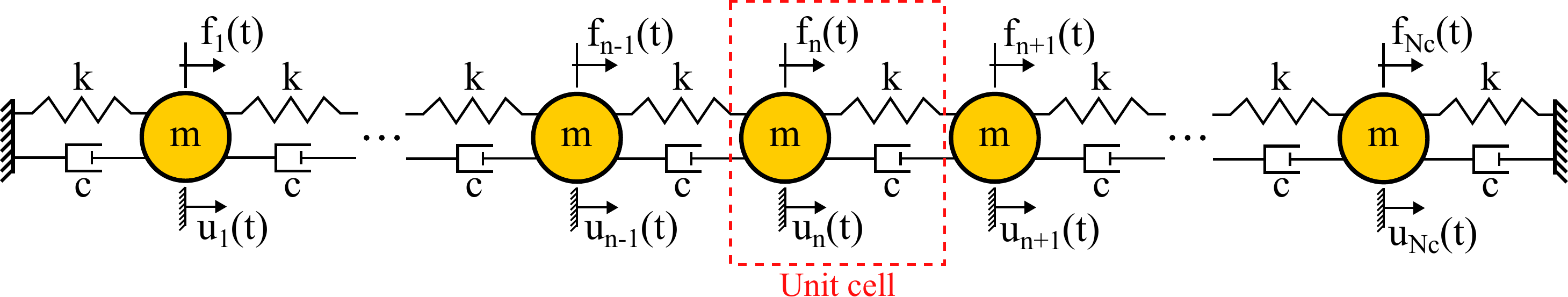}
\caption{One-dimensional lumped system with spatially periodic feedback.}
\label{fig: feedback_system}
\end{figure*}

The spatially periodic feedback system consists of a chain of identical unit cells, each comprising a single-degree-of-freedom (DOF) mass-spring-damper assembly with constant properties, as depicted in Fig.~\ref{fig: feedback_system}. Each mass $m$ is connected to its neighbors by springs of stiffness $k$ and dampers with a viscous damping coefficient $c$. In addition, each unit cell incorporates an active feedback mechanism applied directly to the mass.


The general equation of motion governing the displacement of the mass in the $n$-th unit cell, $u_n(t)$ 
can be written
\begin{align}
    \label{eq: feedback_motion}
    m\ddot{u}_n(t) = k\bigl(u_{n+1}(t) - u_n(t)\bigr) + c\bigl(\dot{u}_{n+1}(t) - \dot{u}_n(t)\bigr)\notag \\- k\bigl(u_n(t) - u_{n-1}(t)\bigr) - c\bigl(\dot{u}_n(t) - \dot{u}_{n-1}(t)\bigr) + f_n(t),
\end{align}
where $\dot{u}_n(t)$ and $\ddot{u}_n(t)$ denote the first and second time derivatives of $u_n(t)$, respectively, and $f_n(t)$ is the active feedback force applied to the $n$-th mass. Adopting a generalized feedback law
\begin{align}\label{eq: feedback_force}
    f_n(t) &= g^a \Bigl( \ddot{u}_n(t) - \ddot{u}_{n-1}(t) \Bigr) +  g^v \Bigl( \dot{u}_n(t) - \dot{u}_{n-1}(t) \Bigr)\notag\\& + g^d \Bigl( u_n(t) - u_{n-1}(t) \Bigr).
\end{align}

 We hereafter omit the explicit time dependence for brevity.
To derive the dispersion relation, 
the Floquet-Bloch theorem is invoked, whereby the response of the periodic system satisfies $ u_{n+1}(t) = e^{-j\mu} u_n(t)$, $u_{n-1}(t) = e^{j\mu} u_n(t)$. 
%
Substituting those expressions 
into Eq.~(\ref{eq: feedback_motion}), and  Eq.~(\ref{eq: feedback_force}), 
and assuming a time-harmonic response $u_n(t) = \hat{u}_n(\omega)\,e^{j\omega t}$, the following characteristic equation is obtained for the spatially periodic feedback system
\begin{align}\label{eq: dispersion_relation_normalized}
   & \Bigl[ \gamma^a \omega_0^2 (1 - e^{j\mu}) - 1 \Bigr] \Omega^2\\+ 
   & j \left[ \frac{b}{\sqrt{mk}} (2 - e^{-j\mu} - e^{j\mu}) + \gamma^v \omega_0 (e^{j\mu} - 1) \right] \Omega \notag\\+
   & \Bigl[ 2 - e^{-j\mu} - e^{j\mu} + \gamma^d (e^{j\mu} - 1) \Bigr] = 0,\notag
\end{align}
 where we used the normalized frequency $\Omega = \omega / \omega_0$, with 
$\omega_0^2 = k/m$, 
and the normalized gain ratios $\gamma^\Delta = g^\Delta / k$, $\Delta \in \{a, v, d\}$.  
For each fixed dimensionless wavenumber $\mu$, Eq.~(\ref{eq: dispersion_relation_normalized}) defines a quadratic eigenvalue problem in the normalized frequency $\Omega$. Solving it yields 
the dispersion diagram of the system.
The dispersion diagram is analyzed over the range $\mu \in [-\pi, \pi]$, which corresponds to the First Brillouin Zone (FBZ). 

\subsubsection{Stability analysis and time-domain response}

For the stability analysis and time-domain response, the finite structure depicted in Fig.~\ref{fig: feedback_system} is considered. 
We obtain a finite structure by truncating the
infinite system to an odd number of unit cells $N_c = 2N + 1$, with $N \in \mathbb{N}$, so we can excite the structure in a central mass. We consider a disturbance $w_n(t)$ at any instant $t$, and conveniently adopt fixed boundary conditions at both ends, to disregard the zero-frequency vibration modes. 
The equation of motion for all the $N_c$ unit cells, with $f_1(t) = 0$ for the first cell of the structure, can be written as a system of $N_c$ coupled second-order ordinary differential equations in matrix form
\begin{align}\label{eq: finite_system_motion_general}
    \underbrace{([M] + [G^a])}_{\bar{M}}  \{\ddot{u}(t)\} + \underbrace{([C] + [G^v])}_{\bar{C}} \{\dot{u}(t)\} \notag\\+ \underbrace{([K] + [G^d])}_{\bar{K}} \{u(t)\} = \{w(t)\},
\end{align}
where $[M]$, $[C]$, and $[K]$ are the mass, damping, and stiffness matrices of the passive system, respectively, while $[G^a]$, $[G^v]$, and $[G^d]$ are the gain matrices associated with the feedback action. The vectors $\{u(t)\}$, $\{\dot{u}(t)\}$, and $\{\ddot{u}(t)\}$ denote the displacement, velocity, and acceleration of the masses, respectively, and $\{w(t)\}$ is the external perturbation vector applied to the masses. These matrices can be expressed in terms of the system parameters as
\begin{align}
    [M] &= m [I]_{N_c}, \qquad [C] = c [D_1]_{N_c}, \notag\\ [K] &= k [D_1]_{N_c}, \qquad [G^\Delta]=g^\Delta[D_2]_{N_c},
\end{align}
where $[D_1]_{N_c}$ and $[D_2]_{N_c}$ are given by
\begin{align}
    [D_1]_{N_c} = 
    \begin{bmatrix}
        2 & -1 & 0 & \cdots & 0 & 0\\
        -1 & 2 & -1 & \cdots & 0 & 0 \\
        0 & -1 & 2 & \cdots & 0 & 0 \\
        \vdots & \vdots & \vdots & \ddots & \vdots & \vdots \\
        0 & 0 & 0 & \cdots & 2 & -1 \\
        0 & 0 & 0 & \cdots & -1 & 2
    \end{bmatrix}_{N_c},\\
    [D_2]_{N_c} =
    \begin{bmatrix}
        0 & 0 & 0 & \cdots & 0 & 0 \\
        1 & -1 & 0 & \cdots & 0 & 0 \\
        0 & 1 & -1 & \cdots & 0 & 0 \\
        \vdots & \vdots & \vdots & \ddots & \vdots & \vdots \\
        0 & 0 & 0 & \cdots & 1 & -1
    \end{bmatrix}_{N_c}.
\end{align}

This system of coupled differential equations can be recast as a first-order linear system through a state-space representation. To this end, the state vector is defined as $\{x(t)\} = \{u(t),\, \dot{u}(t)\}^T$. The state-space representation of the closed-loop system, with explicit time dependence omitted for brevity, is then written as
\begin{align}\label{eq: state_space_representation}
    \begin{Bmatrix}
        \dot{x}
    \end{Bmatrix} =
    \underbrace{
    \begin{bmatrix}
        [0]_{N_c} & [I]_{N_c} \\
        -\bar{M}^{-1} \bar{K} & -\bar{M}^{-1} \bar{C}
    \end{bmatrix}
    }_A
    \begin{bmatrix}
        x
    \end{bmatrix} 
    +
    \underbrace{    
    \begin{bmatrix}
        [0]_{N_c} \\
        \bar{M}^{-1}
    \end{bmatrix}
    }_B
    \{w\}.
\end{align}

The stability of a system in state-space form can then be assessed through the eigenvalues $\lambda$ of the state matrix $A$. The system is asymptotically stable if and only if all eigenvalues of $A$ have strictly negative real parts, i.e., $\max\bigl(\text{Re}(\lambda)\bigr) < 0$
. The stability analysis is therefore carried out by computing the eigenvalues of $A$ for varying feedback gain parameters $g^a$, $g^v$, and $g^d$. To examine the time-domain response, the finite system is subjected to a localized sine burst transient excitation at the central unit cell ($n = N + 1$), with the carrier frequency, modulated by a Hanning window, strategically chosen to highlight the effects observed in the dispersion diagram. The response is obtained by numerically integrating Eq.~(\ref{eq: state_space_representation}). 
The time-domain analysis enables verification of directional amplification or attenuation effects within specific frequency ranges, which constitute clear indicators of non-reciprocal wave propagation.

\subsection{Space-time modulation-induced non-reciprocity}

\begin{figure*}[t]
\centering
\includegraphics[width=0.9\textwidth]{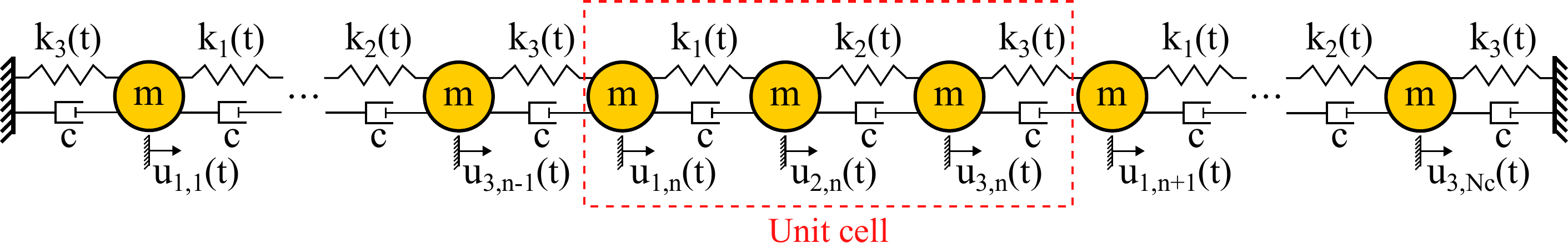}
\caption{One-dimensional lumped system with space-time modulated stiffness.}
\label{fig: stm_system}
\end{figure*}

For the space-time modulation case, a one-dimensional chain of periodic mass-spring-damper unit cells is considered. In this case, each unit cell contains $R = 3$ DOF, and the springs' stiffness is modulated periodically in space and time as depicted in Fig.~\ref{fig: stm_system}. Each unit cell consists of three equal masses $m$ connected by equal dampers with a viscous damping coefficient $c$. The stiffness of the springs connecting the masses is modulated in space and time according to
\begin{align}\label{eq: stm_modulation}
    k_r(t) = k_0 \Bigl[ 1 + \alpha_m \Theta(r,t) \Bigr],
\end{align}
where $k_0$ is the average stiffness, $\alpha_m$ is the modulation amplitude, and $\Theta(r,t)$ is a time-periodic function that defines the modulation pattern. The equation of motion governing the displacement of the $R$ masses in the $n$-th unit cell $\{u\}_n(t)$ 
at any instant $t$ can be written in matrix form, omitting the explicit time dependence of the displacements and stiffness for brevity, as Eqn.~\eqref{eq: stm_motion}.

\begin{figure*}[t]
    \begin{align}\label{eq: stm_motion}
\begin{split}
    &
    \begin{bmatrix}
        m & 0 & 0 \\
        0 & m & 0 \\
        0 & 0 & m
    \end{bmatrix}
    \{\ddot{u}\}_n
    +
    \begin{bmatrix}
        2c & -c & 0 \\
        -c & 2c & -c \\
        0  & -c & 2c
    \end{bmatrix}
    \{\dot{u}\}_n+
    \begin{bmatrix}
        0 & 0 & -c \\
        0 & 0 &  0 \\
        0 & 0 &  0
    \end{bmatrix}
    \{\dot{u}\}_{n-1}
    +
    \begin{bmatrix}
        0 & 0 &  0 \\
        0 & 0 &  0 \\
       -c & 0 &  0
    \end{bmatrix}
    \{\dot{u}\}_{n+1} +
    \begin{bmatrix}
        k_1 + k_3 & -k_1         & 0         \\
        -k_1      & k_1 + k_2    & -k_2      \\
        0         & -k_2         & k_2 + k_3
    \end{bmatrix}
    \{u\}_n\\&
    +
    \begin{bmatrix}
        0 & 0 & -k_3 \\
        0 & 0 &  0   \\
        0 & 0 &  0
    \end{bmatrix}
    \{u\}_{n-1}
    +
    \begin{bmatrix}
        0    & 0 & 0 \\
        0    & 0 & 0 \\
       -k_3  & 0 & 0
    \end{bmatrix}
    \{u\}_{n+1}
    = 
    \{0\}
\end{split}
\end{align}
\end{figure*}
where $\{u\}_n(t) = \{u_1(t), u_2(t), u_3(t)\}^T$ is the displacement vector of the three masses in the $n$-th unit cell and $\{\dot{u}\}_n(t)$, $\{\ddot{u}\}_n(t)$ are the corresponding velocity and acceleration vectors, respectively. 
As in the feedback case, the dispersion relation is obtained 
invoking the Floquet-Bloch theorem, resulting in Eqn.~\eqref{eq: stm_motion_bloch_floquet}.
\begin{figure*}[t]
    \begin{align}\label{eq: stm_motion_bloch_floquet}
\begin{split}
    &
    \underbrace{
    \begin{bmatrix}
        m & 0 & 0 \\
        0 & m & 0 \\
        0 & 0 & m
    \end{bmatrix}
    }_{[M]}
    \{\ddot{u}\}_n(t)
    +
    \underbrace{
    \begin{bmatrix}
        2c & -c & -ce^{j\mu} \\
        -c & 2c & -c \\
        -c e^{-j\mu}  & -c & 2c
    \end{bmatrix}
    }_{[C(\mu)]}
    \{\dot{u}\}_n(t)
    +
    \underbrace{
    \begin{bmatrix}
        k_1 + k_3 & -k_1         & -k_3 e^{j\mu}         \\
        -k_1      & k_1 + k_2    & -k_2      \\
        -k_3 e^{-j\mu}         & -k_2         & k_2 + k_3
    \end{bmatrix}
    }_{[K(\mu,t)]}
    \{u\}_n(t)
    = \{0\}
\end{split}
\end{align}
\end{figure*}

The spatial periodicity of the system is reflected in the dependence of the damping and stiffness matrices on the wavenumber $\mu$, while the time periodicity is manifested in the stiffness matrix $[K(\mu,t)]$ due to the space-time modulation of the spring stiffness, $[K(\mu,t + T_m)] = [K(\mu,t)]$, for all $t$, where $T_m = \frac{2\pi}{\omega_m}$ is the modulation period and $\omega_m$ is the modulation frequency. Since the stiffness matrix $[K(\mu,t)]$ is time-periodic, it can be expressed as a Fourier series
\begin{align}\label{eq: fourier_series_stiffness}
    [K(\mu,t)] = \sum_{p=-\infty}^{\infty} [\hat{K}_p(\mu)] e^{jp\omega_m t},
\end{align}
where $[\hat{K}_p(\mu)]$ are the Fourier coefficients of the stiffness matrix. By exploiting the system's space-time periodicity, the Plane Wave Expansion (PWE) method can be applied in conjunction with the Floquet-Bloch theorem. Specifically, the double periodicity allows the displacement response of the $n$-th unit cell to be represented as a superposition of plane waves in both space and time using Fourier series expansion
\begin{align}\label{eq: pwe_response}
    \{u\}_n(t) = \sum_{q=-\infty}^{\infty} \{\hat{u}\}_q(\mu)\, e^{i\bigl(n\mu + (\omega + q\omega_m)t\bigr)},
\end{align}
where $\{\hat{u}\}_q(\mu)$ are the complex amplitude vectors associated with the $q$-th harmonic, $\mu$ is the dimensionless wavenumber, $\omega$ is the base frequency, and $q\omega_m$ are the frequency harmonics generated by the space-time modulation. Substituting Eq.~(\ref{eq: pwe_response}) and Eq.~(\ref{eq: fourier_series_stiffness}) into Eq.~(\ref{eq: stm_motion_bloch_floquet}) and truncating the Fourier series to a finite number of harmonics $|q| \leq Q$ and 
$|p| \leq P$, we obtain
\begin{align}\label{eq: pwe_matrix_align}
    &- \sum_{q = -Q}^{Q} (\omega + q\omega_m)^2 [M] \{\hat{u}\}_q(\mu)\, e^{i\bigl(n\mu + (\omega + q\omega_m)t\bigr)}\\& + \sum_{q = -Q}^{Q} j(\omega + q\omega_m) [C(\mu)] \{\hat{u}\}_q(\mu)\, e^{i\bigl(n\mu + (\omega + q\omega_m)t\bigr)} \notag\\
    & + \sum_{p=-P}^{P} \sum_{q=-Q}^{Q} [\hat{K}_p(\mu)] \{\hat{u}\}_q(\mu)\, e^{i\bigl(n\mu + (\omega + (q+p)\omega_m)t\bigr)} = \{0\}.\notag
\end{align}

Substituting $q = \bar{q} - p$ in the third term of Eq.~(\ref{eq: pwe_matrix_align}) and, without loss of generality, setting $q = \bar{q}$ in the first and second terms, all the terms can be expressed in terms of the same exponential function $e^{i\bigl(n\mu + (\omega + \bar{q}\omega_m)t\bigr)}$, which allows the application of the orthogonality property of the exponential functions. This procedure implies that to satisfy Eq.~(\ref{eq: pwe_matrix_align}), we have for each $\bar{q}$
\begin{align}\label{eq: pwe_eigenvalue_problem_normalized}
    &\Bigl[ (\Omega + \bar{q}\Omega_m)^2 [I]_{3} + j(\Omega + \bar{q}\Omega_m) \frac{1}{\sqrt{mk_0}} [C(\mu)] \Bigr] \{\hat{u}\}_{\bar{q}}(\mu)\notag\\& + \sum_{p=-P}^{P} \frac{1}{m \omega_0^2} [\hat{K}_p(\mu)]  \{\hat{u}\}_{\bar{q}-p}(\mu) = \{0\},
\end{align}
where we normalized the equation by the modulation frequency $\Omega_m = \omega_m / \omega_0$. We propose a harmonic modulation of the springs' stiffness as
\begin{align}\label{eq: harmonic_modulation}
    \Theta(r,t) = \cos\bigl(\omega_m t - \phi_r\bigr) \quad \text{with} \quad \phi_r = \frac{2\pi r}{R}.
\end{align}

Substituting Eq.~(\ref{eq: harmonic_modulation}) into the expression for $[K(\mu,t)]$ implies that its Fourier coefficients are nonzero only for $p = 0, \pm 1$. Therefore, the response of the system can be approximated by considering only the first-order harmonics ($Q = 1$), and the Fourier series expansion of the stiffness matrix can be truncated to three terms ($P = 1$). Under these assumptions, the complex amplitude vectors $\{\hat{u}\}_{\bar{q}}(\mu)$ are nonzero only for $\bar{q} = 0, \pm Q$, which implies that the term $\{\hat{u}\}_{\bar{q}-p}(\mu)$ in Eq.~(\ref{eq: pwe_eigenvalue_problem_normalized}) is nonzero only for $\bar{q} - Q \leq p \leq \bar{q} + Q$. Applying these conditions to Eq.~(\ref{eq: pwe_eigenvalue_problem_normalized}) yields
\begin{align*}
    &\left(-\begin{Bmatrix}
        \{(\Omega + \bar{q}\Omega_m)^2\}_{\bar{q}=-1} \\
        \{(\Omega + \bar{q}\Omega_m)^2\}_{\bar{q}=0} \\
        \{(\Omega + \bar{q}\Omega_m)^2\}_{\bar{q}=1}
    \end{Bmatrix}
    \begin{bmatrix}
        \mathbf{I}_{3} & \mathbf{0}_{3} & \mathbf{0}_{3} \\
        \mathbf{0}_{3} & \mathbf{I}_{3} & \mathbf{0}_{3} \\
        \mathbf{0}_{3} & \mathbf{0}_{3} & \mathbf{I}_{3}
    \end{bmatrix}
     \right.\\
    &\left.+ \frac{j}{\sqrt{mk}} 
    \begin{Bmatrix}
        \{(\Omega + \bar{q}\Omega_m)\}_{\bar{q}=-1} \\
        \{(\Omega + \bar{q}\Omega_m)\}_{\bar{q}=0} \\
        \{(\Omega + \bar{q}\Omega_m)\}_{\bar{q}=1}
    \end{Bmatrix}
    \begin{bmatrix}
        \mathbf{C} & \mathbf{0}_{3} & \mathbf{0}_{3} \\
        \mathbf{0}_{3} & \mathbf{C} & \mathbf{0}_{3} \\
        \mathbf{0}_{3} & \mathbf{0}_{3} & \mathbf{C}
    \end{bmatrix}
 \right.\\
    &\left.+ \frac{1}{m \omega_0^2}
    \begin{bmatrix}
        \mathbf{\hat{K}_{p = 0}} & \mathbf{\hat{K}_{p = -1}} & \mathbf{\hat{K}_{p = -2}} \\
        \mathbf{\hat{K}_{p = 1}} & \mathbf{\hat{K}_{p = 0}} & \mathbf{\hat{K}_{p = -1}} \\
        \mathbf{\hat{K}_{p = 2}} & \mathbf{\hat{K}_{p = 1}} & \mathbf{\hat{K}_{p = 0}} \\
    \end{bmatrix}
    \right)
    \begin{Bmatrix}
        \{\hat{u}\}_{q=-1} \\
        \{\hat{u}\}_{q=0} \\
        \{\hat{u}\}_{q=1}
    \end{Bmatrix}
    = \{0\},
\end{align*}
a quadratic eigenvalue problem for the normalized frequency $\Omega$ as a function of the dimensionless wavenumber $\mu$. The solution to this eigenvalue problem yields the system's dispersion diagram, which can be analyzed to identify non-reciprocal wave propagation. 
As in the feedback case, the analysis of the dispersion diagram is carried out over the range $\mu \in [-\pi, \pi]$, which corresponds to the FBZ. 
The space-time modulation 
induces coupling between different harmonics of the response: 
the dispersion relation of the system implies that, for each wavenumber $\mu$, there are $R$ groups of $2Q + 1$ eigenvalues. Each group comprises the central frequency $\Omega_r$ and its associated harmonics $\Omega_r + q\Omega_m$, where $q = -Q, \ldots, Q$. Then, obtaining the solution introduces the challenge of correctly identifying, for each wavenumber $\mu$, the $R$ branches of the dispersion diagram that correspond to the desired plane wave response. To this end, the procedure proposed by~\cite{VILA2017363} is adopted, in which the magnitude of the complex amplitude vectors $\{\hat{u}\}_q(\mu)$ is used as a criterion to identify the corresponding branches of the dispersion diagram associated with the plane wave response. 

\subsubsection{Stability analysis and time-domain response}

The stability analysis and time-domain response of the space-time modulated system are carried out by considering a finite structure composed of $N_c$ unit cells, with fixed boundary conditions at both ends. The equations of motion for the finite system can be derived in the state space representation, with $\{x(t)\} = \{u(t),\, \dot{u}(t)\}^T$, as 
\begin{align}\label{eq: stm_state_space_representation}
 \begin{Bmatrix}
        \dot{x}
    \end{Bmatrix}&=
    \underbrace{
    \begin{bmatrix}
        [0]_{R \cdot N_c} & [I]_{R \cdot N_c} \\
        -[M]^{-1} [K(t)] & -[M]^{-1} [C]
    \end{bmatrix}
    }_{[A(t)]}
    \begin{Bmatrix}
        x
    \end{Bmatrix} 
   \\& +
    \underbrace{    
    \begin{bmatrix}
        [0]_{R \cdot N_c} \\
        [M]^{-1}
    \end{bmatrix}
    }_{[B]}
    \{w\},\notag
\end{align}
where $[M]$, $[C]$, and $[K(t)]$ are the mass, damping, and time-varying stiffness matrices of the finite system, respectively. Consider $\{w(t)\} = \{0\}$ 
and initial condition $\{x(t_0)\}$. 
The solution is parametrized by the state transition matrix $[\Phi](t, t_0)$, such that $\{x(t)\} = [\Phi(t, t_0)] \{x(t_0)\}$, or by the fundamental matrix $[\Psi(t)]$, with 
%
$[\Phi(t, t_0)] = [\Psi(t)] [\Psi(t_0)]^{-1}$, and 
$[\dot \Psi(t)]  = [A(t)][\Psi(t)]$. Assuming  
$t_0 = 0$, the $[\Psi(t)]$ matrix can be chosen in the normalized form satisfying $[\Psi(0)] = [I]_{2RN_c}$. 
Due to periodicity, 
$[\Psi(t + T_m)] = [\Psi(t)] [\Psi(T_m)]$. 
This implies that the state transition matrix also satisfies $[\Phi(t + T_m, 0)] = [\Phi(t, 0)] [\Phi(T_m, 0)]$. 
 According to Floquet theory, the stability of a linear periodic system can be determined by computing the monodromy matrix 
  $[\Gamma] = [\Phi(T_m, 0)]$ through numerical integration of $[\dot{\Phi}(t, 0)] = [A(t)] [\Phi(t, 0)]$ over one modulation period $T_m$. The eigenvalues of the monodromy matrix 
 are known as Floquet multipliers; 
 the system is stable if and only if all Floquet multipliers lie within the unit circle in the complex plane \cite{Richards1983}. 
 
 In Sec.~\ref{sec:results_stability}\ , we compute the monodromy matrix for varying modulation parameters $\alpha_m$ and $\Omega_m$ and analyze the corresponding Floquet multipliers. 
As in the feedback case, we examine the time-domain response of the finite system subjected to a localized sine burst transient excitation at the central mass, with the carrier frequency, modulated by a Hanning window, strategically chosen to highlight the effects observed in the dispersion diagram. The response is obtained by numerically integrating Eq.~(\ref{eq: stm_state_space_representation}) using an appropriate time-integration method for time-varying systems. 

\section{Results}

The results for stability, dispersion diagrams, and time-domain responses are presented for both the feedback and space-time modulated systems. They are analyzed with respect to the effects of feedback and modulation parameters on non-reciprocal wave propagation. Stable regions in parameter space are identified, followed by the dispersion diagrams and, finally, the time-domain responses used to validate the theoretical predictions.\

\subsection{Stability analysis}\label{sec:results_stability}

\begin{figure*}[h!]
    \centering
    \begin{subfigure}{0.33\textwidth}
        \centering
        \includegraphics[width=\textwidth]{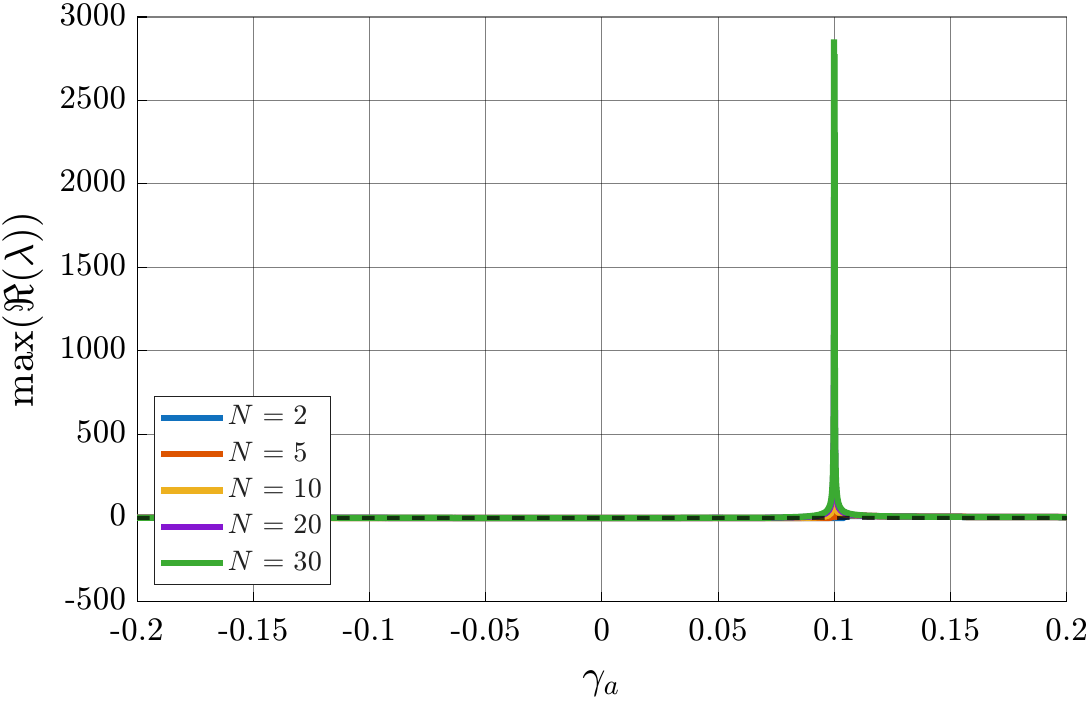}
        \caption{Varying $\gamma^a$}
        \label{fig: feedback_stability_ga}
    \end{subfigure}\hfill
    \begin{subfigure}{0.33\textwidth}
        \centering
        \includegraphics[width=\textwidth]{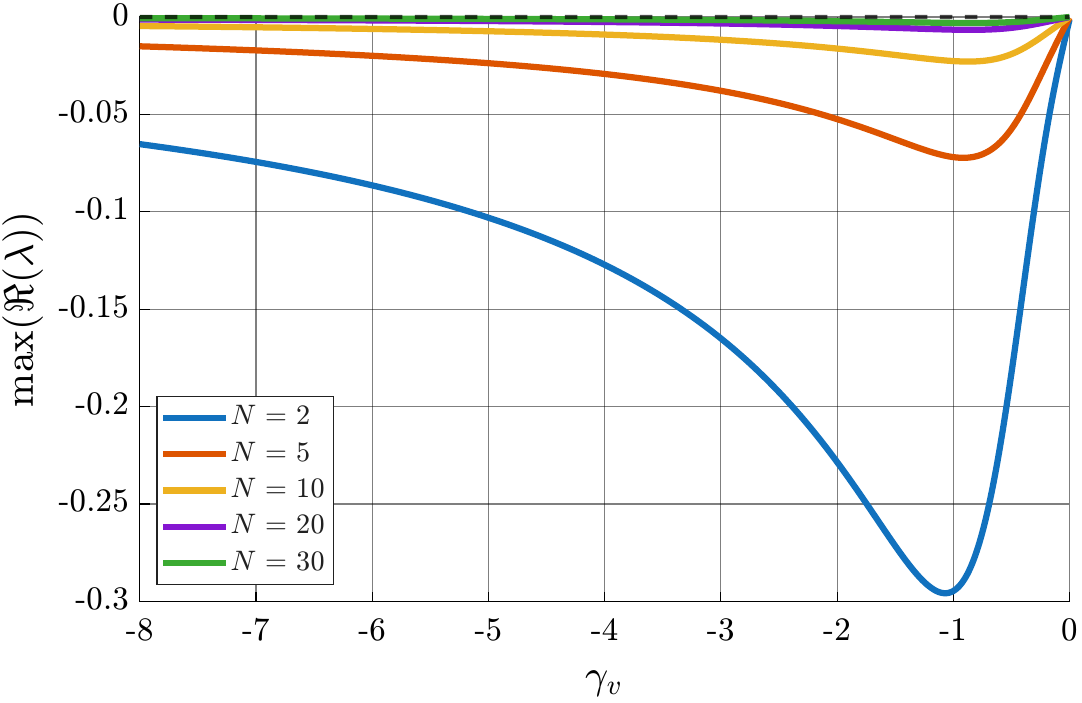}
        \caption{Varying $\gamma^v$}
        \label{fig: feedback_stability_gv}
    \end{subfigure}
    \begin{subfigure}{0.33\textwidth}
        \centering
        \includegraphics[width=\textwidth]{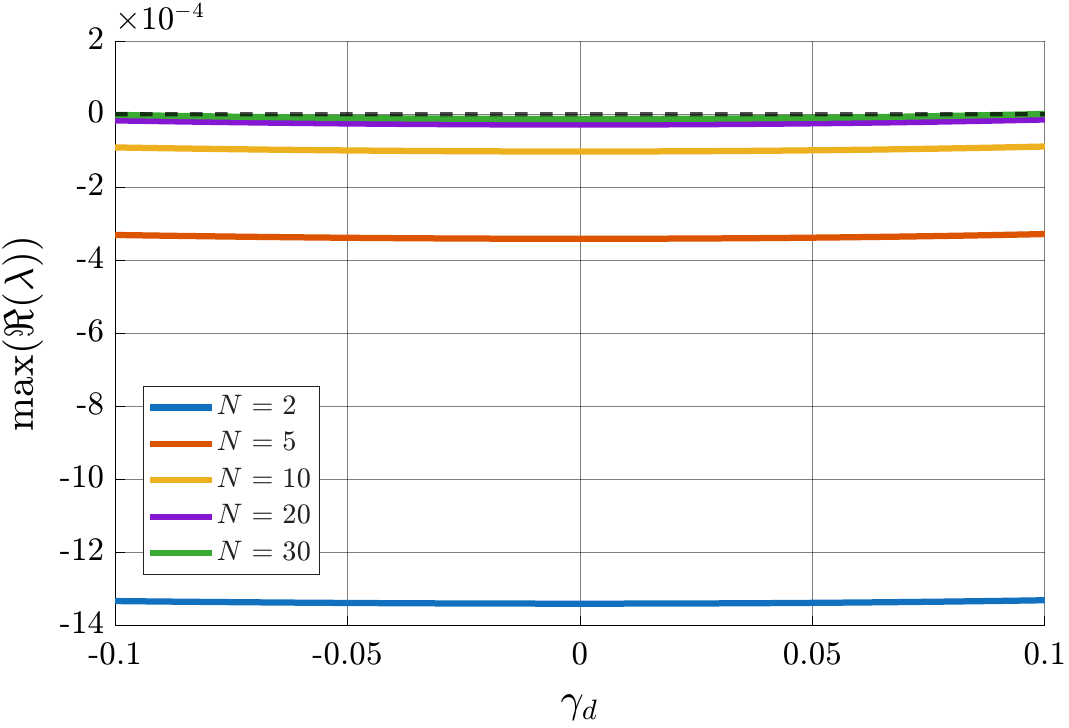}
        \caption{Varying $\gamma^d$}
        \label{fig: feedback_stability_gd}
    \end{subfigure}
    \caption{Stability analysis of the feedback system: maximum real part of the eigenvalues of the state matrix $A$ for varying feedback gain parameters.}
    \label{fig: feedback_stability}
\end{figure*}

\begin{figure*}[H]
    \centering
    \begin{subfigure}{0.48\textwidth}
        \centering
        \includegraphics[width=\textwidth]{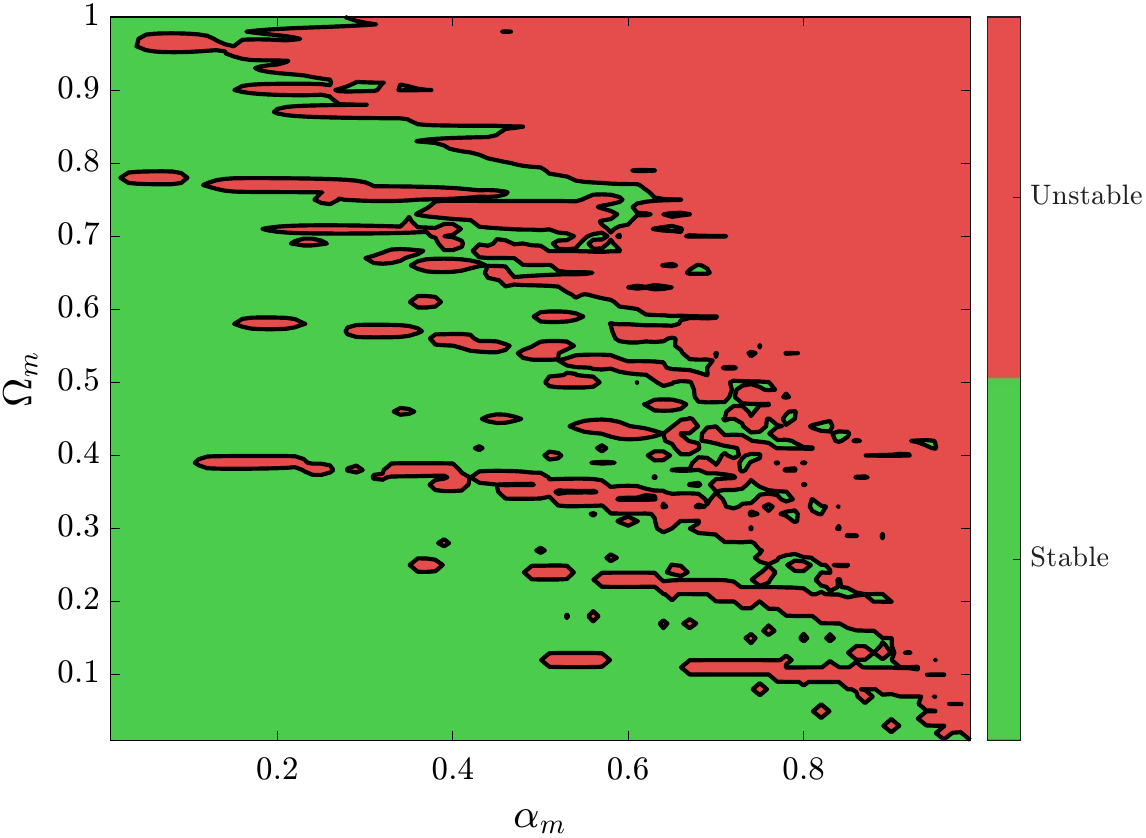}
        \caption{Stability region for $N = 2$}
        \label{fig: stm_stability_contour}
    \end{subfigure}\hfill
    \begin{subfigure}{0.48\textwidth}
        \centering
        \includegraphics[width=\textwidth]{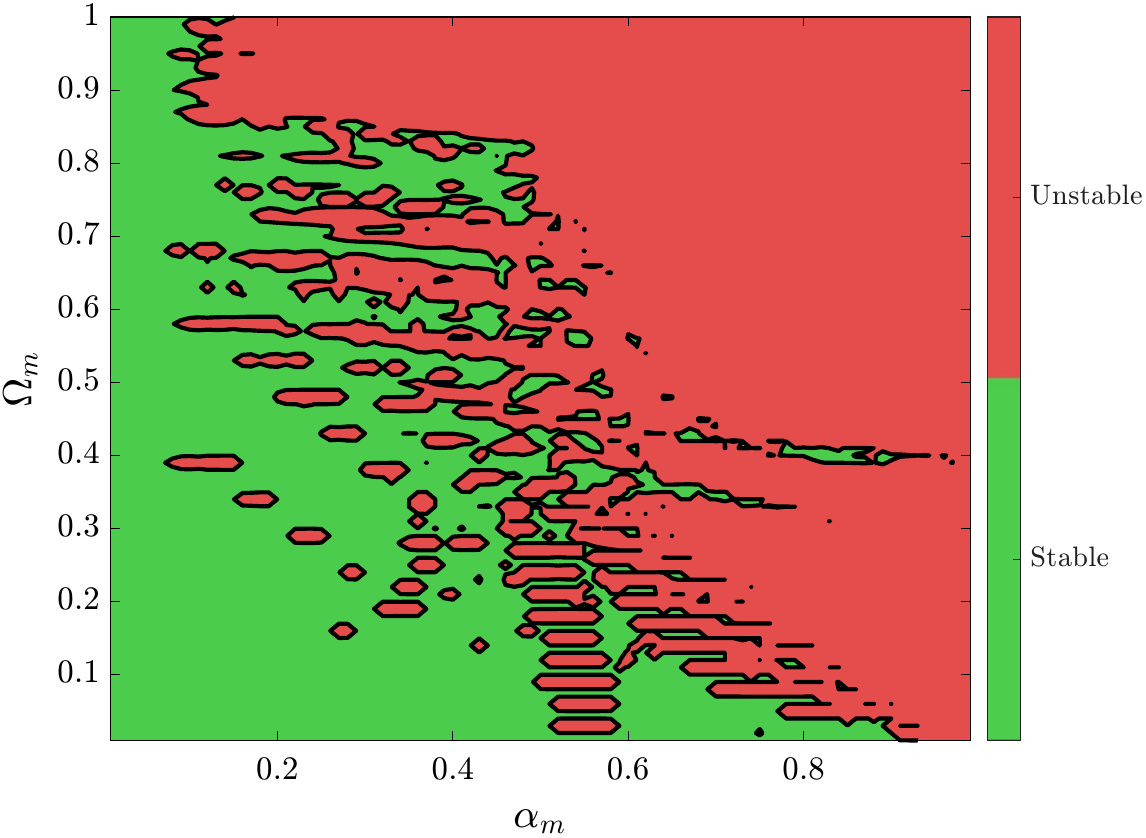}
        \caption{Stability region for $N = 10$}
        \label{fig: stm_stability_map}
    \end{subfigure}
    \caption{Stability/instability regions of the space-time modulated system in the $(\alpha_m,\, \Omega_m)$ parameter space for $N = 2$ and $N = 10$.}
    \label{fig: stm_stability}
\end{figure*}

To assess the stability of the feedback and space-time modulated systems, the eigenvalues of the state matrix $A$ for the feedback case and the Floquet multipliers of the monodromy matrix $[\Gamma]$ for the space-time modulated case are computed for varying feedback gain parameters $\gamma^a$, $\gamma^v$, and $\gamma^d$, and modulation parameters $\alpha_m$ and $\Omega_m$, respectively. For the STM system, the stability analysis is carried out over $\Omega_m \in [0.01,\,1]$ and $\alpha_m \in [0,\,0.99]$, both with a step size of $0.01$, for two values of $N$ ($N = 2$ and $N = 10$) to examine the effect of the number of unit cells on system stability. Fig.~\ref{fig: feedback_stability} shows the maximum real part of the eigenvalues of the state matrix $A$ as a function of the feedback gain parameters. For STM, because direct value maps are not sufficiently clear for interpretation, only stability/instability regions are presented in Fig.~\ref{fig: stm_stability}. Stable regions in parameter space correspond to points where all Floquet multipliers lie within the unit circle.

Fig.~\ref{fig: feedback_stability} shows that the stability of the feedback system is highly sensitive to the feedback gain parameters and depends on the type of feedback. For acceleration feedback gain $g^a$, no stable region exists within $\gamma^a \in [-0.2,\, 0.2]$. Moreover, a singular value can be identified at $\gamma^a \approx 0.1$, at which the feedback interconnection is not well-posed. 
However, for the displacement feedback gain $g^d$, the system's stability depends on the number of unit cells $N_c$. With $N_c \geq 30$ the system is stable only for $|\gamma^d| < 0.1$. 
For the velocity feedback gain $g^v$, the feedback produces a damping effect, which
stabilizes the system for $\gamma^v < 0$. 
For this reason, we focus on the velocity feedback gain in computing the dispersion diagrams and time-domain response. On the other hand, 
Fig.~\ref{fig: stm_stability} shows that, for the space-time modulated system, 
the stability region shrinks as $N$, $\alpha_m$, or $\Omega_m$ increases.

\subsection{Dispersion diagrams}\label{sec:results_dispersion}

\begin{figure*}
    \centering
    \begin{subfigure}{0.44\textwidth}
        \centering
        \includegraphics[width=\textwidth]{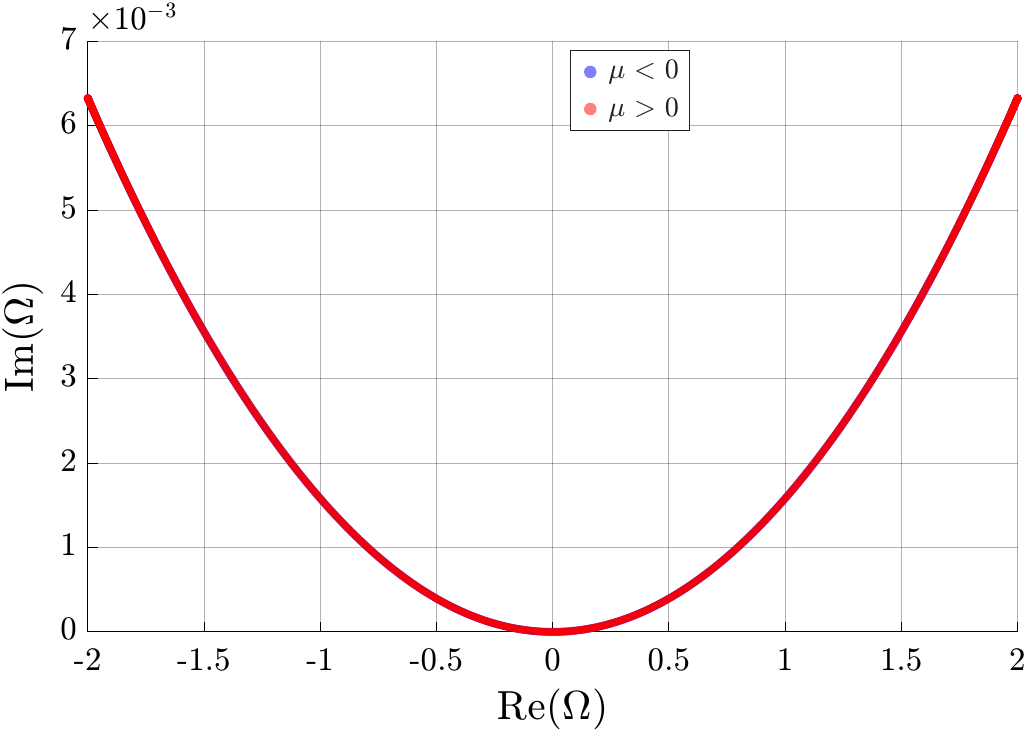}
        \caption{Dispersion diagram for the passive system}
        \label{fig: feedback_dispersion_gv_0}
    \end{subfigure}\hfill
    \begin{subfigure}{0.44\textwidth}
        \centering
        \includegraphics[width=\textwidth]{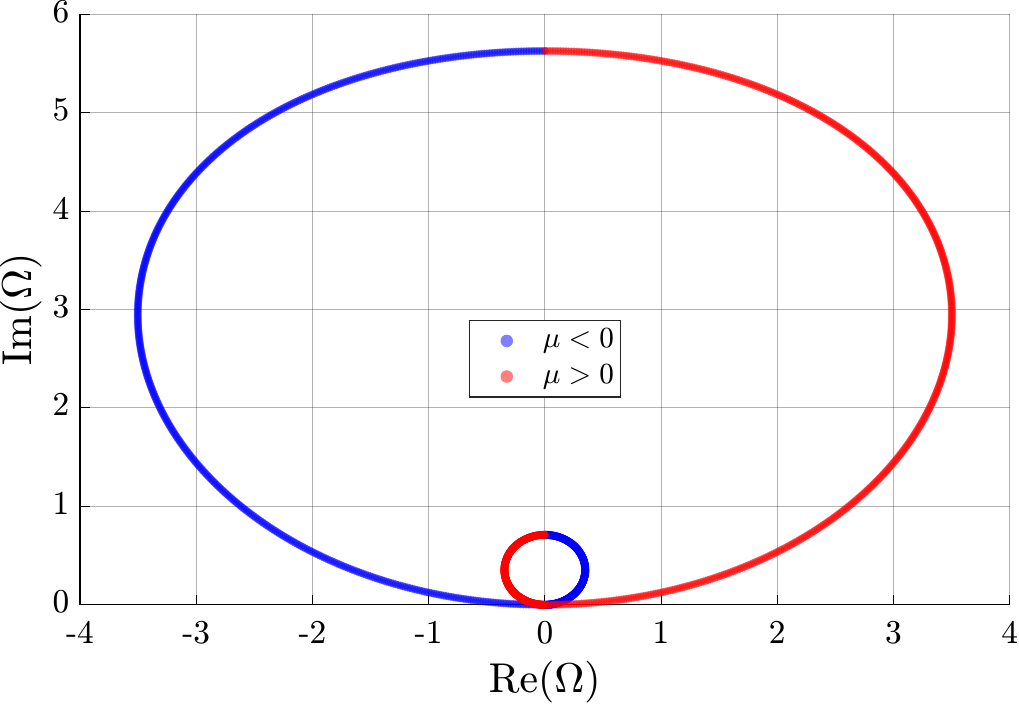}
        \caption{Dispersion diagram for the feedback system with $\gamma^v = -1$}
        \label{fig: feedback_dispersion_gv_1_comparison}
    \end{subfigure}
    \caption{Comparison of dispersion diagrams for the passive system and the feedback system ($\gamma^v = -1$).}
    \label{fig: feedback_dispersion_comparisson}
\end{figure*}

\begin{figure*}
    \centering
    \begin{subfigure}{0.44\textwidth}
        \centering
        \includegraphics[width=\textwidth]{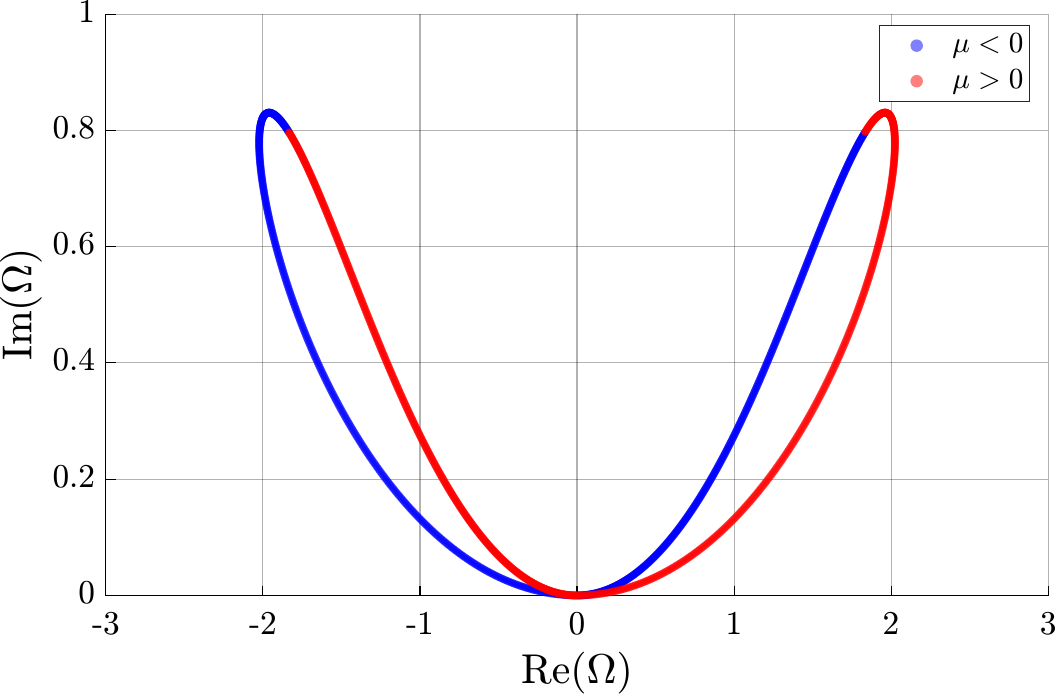}
        \caption{$\gamma^v = -0.25$}
        \label{fig: feedback_dispersion_gv_025}
    \end{subfigure}\hfill
    \begin{subfigure}{0.44\textwidth}
        \centering
        \includegraphics[width=\textwidth]{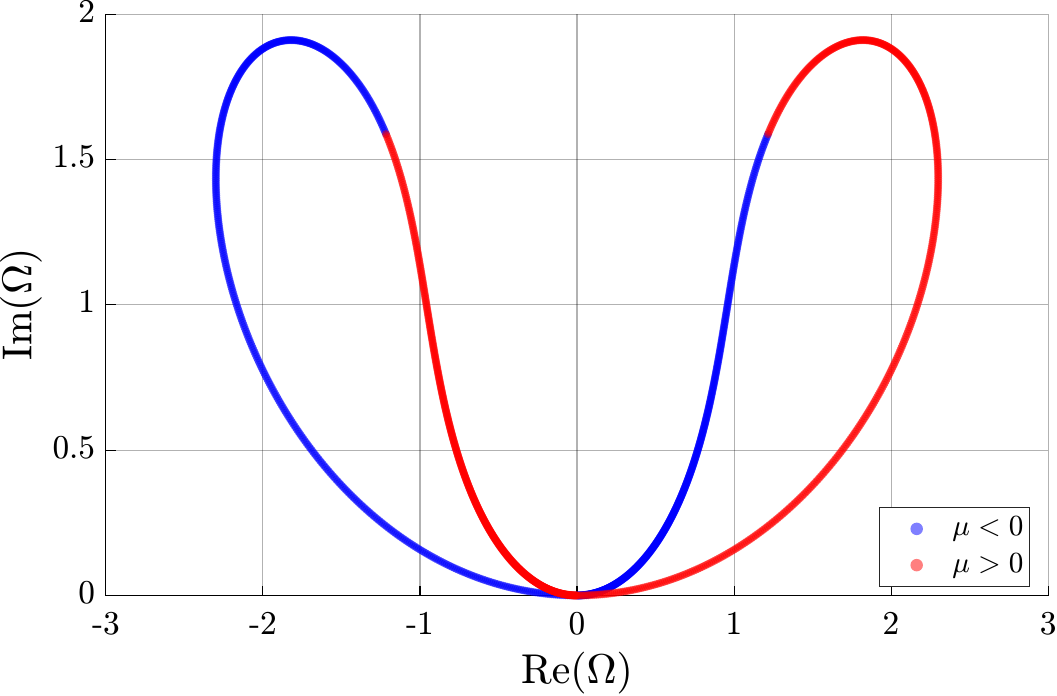}
        \caption{$\gamma^v = -0.5$}
        \label{fig: feedback_dispersion_gv_050}
    \end{subfigure}
    \begin{subfigure}{0.44\textwidth}
        \centering
        \includegraphics[width=\textwidth]{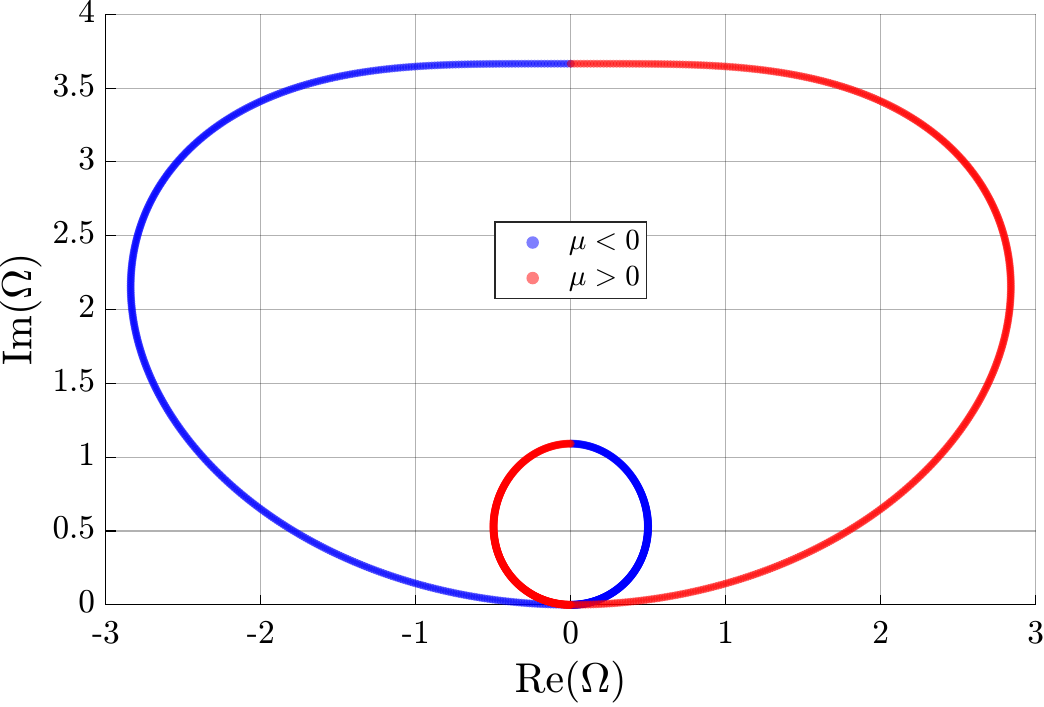}
        \caption{$\gamma^v = -0.75$}
        \label{fig: feedback_dispersion_gv_075}
    \end{subfigure}\hfill
    \begin{subfigure}{0.44\textwidth}
        \centering
        \includegraphics[width=\textwidth]{Figures/Dispersion/dispersion_feedback_complex_gv_-1.pdf}
        \caption{$\gamma^v = -1$}
        \label{fig: feedback_dispersion_gv_1}
    \end{subfigure}
    \begin{subfigure}{0.44\textwidth}
        \centering
        \includegraphics[width=\textwidth]{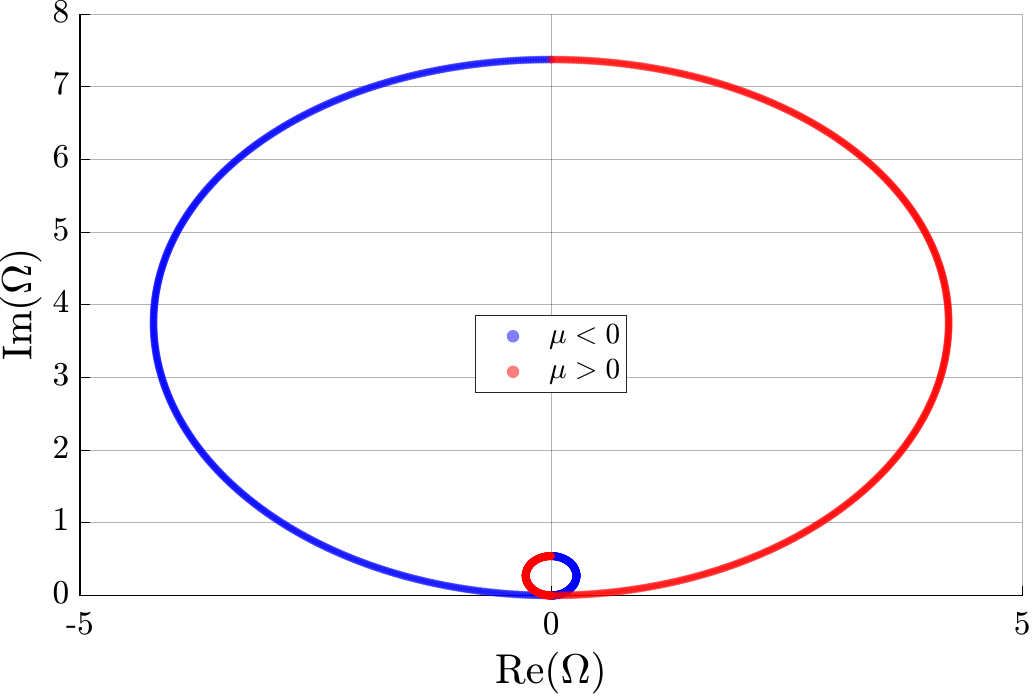}
        \caption{$\gamma^v = -1.25$}
        \label{fig: feedback_dispersion_gv_125}
    \end{subfigure}\hfill
    \begin{subfigure}{0.44\textwidth}
        \centering
        \includegraphics[width=\textwidth]{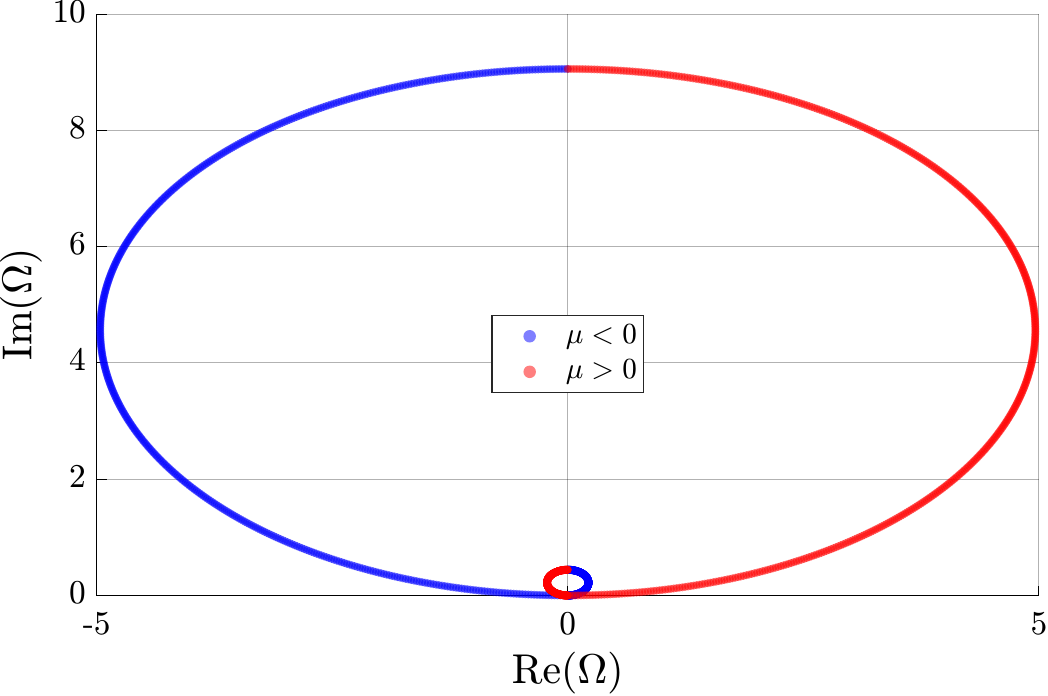}
        \caption{$\gamma^v = -1.5$}
        \label{fig: feedback_dispersion_gv_150}
    \end{subfigure}
    \caption{Projection of dispersion diagrams on complex frequency plane for the feedback system for varying velocity feedback gain $\gamma^v$.}
    \label{fig: feedback_dispersion}
\end{figure*}

Non-reciprocal wave propagation characteristics are expected to manifest as asymmetric dispersion branches and directional band gaps, indicating non-reciprocal behavior. 
Assuming the feedback system, Fig.~\ref{fig: feedback_dispersion_comparisson} shows the difference between the projection of dispersion diagrams onto the complex frequency plane for $\gamma^v = 0$ (passive system) and $\gamma^v = -1$. Figure~\ref{fig: feedback_dispersion} shows the same projection onto the complex frequency plane with the velocity feedback gain varying over $\gamma^v \in [-1.5,\, -0.25]$ in increments of $0.25$. These diagrams show that the velocity feedback gain significantly affects the system's wave-propagation characteristics. As $\gamma^v$ becomes more negative, the dispersion branches become more asymmetric. The imaginary part of the dispersion branches also changes as $\gamma^v$ becomes more negative, increasing in one direction of propagation while decreasing in the other.

The topological interpretation of the dispersion diagram of the feedback system can be obtained from the trajectory of the complex frequency branch $\Omega(\mu)=\Omega_R(\mu)+i\Omega_I(\mu)$ over the FBZ when projected onto the complex plane. When this trajectory forms a closed loop that encloses a nonzero area, the associated winding number is nonzero, indicating a non-trivial non-Hermitian topology. In contrast, open or non-enclosing trajectories are associated with trivial topology (zero winding number). As discussed in related studies, the sign of the winding number is directly linked to the preferred localization direction of skin modes (right- or left-localized)~\cite{Rosa_2020}. Therefore, the increasing asymmetry observed in the complex dispersion branches as $\gamma^v$ becomes more negative is consistent with stronger non-trivial topological behavior and enhanced directional localization tendencies in finite structures.

For the STM system, to better compare $\mathrm{Re}(\Omega)$ and $\mathrm{Im}(\Omega)$, the dispersion analysis is presented for the four combinations defined by $\alpha_m \in \{0.05,\, 0.2\}$ and $\Omega_m \in \{0.05,\, 0.2\}$. The dispersion diagrams reveal that both the modulation amplitude and the modulation frequency significantly affect the system's wave propagation characteristics. In particular, the modulation amplitude $\alpha_m$ directly influences band gap width. By contrast, the modulation frequency $\Omega_m$ significantly affects non-reciprocity by increasing dispersion asymmetry. Directional band gaps are observed for both propagation directions ($\mu > 0$ and $\mu < 0$), with different frequency intervals depending on direction. 

\begin{figure*}
    \centering
    \begin{subfigure}{0.44\textwidth}
        \centering
        \includegraphics[width=\textwidth]{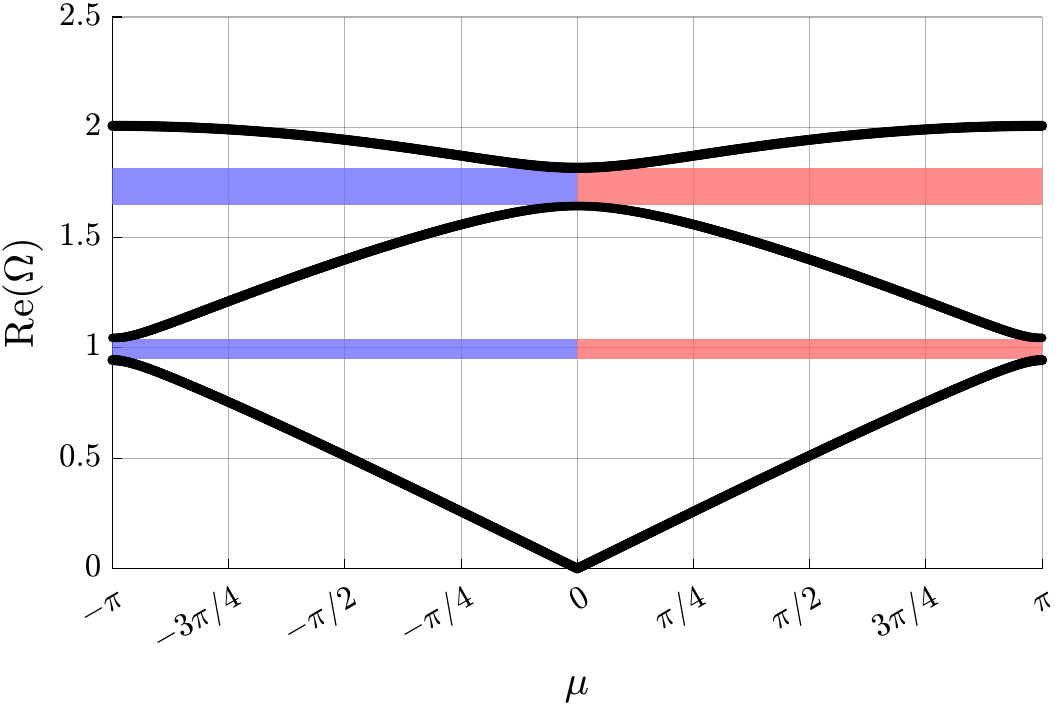}
        \caption{$\mathrm{Re}(\Omega)$ for $\alpha_m = 0.2$, $\Omega_m = 0$}
    \end{subfigure}\hfill
    \begin{subfigure}{0.44\textwidth}
        \centering
        \includegraphics[width=\textwidth]{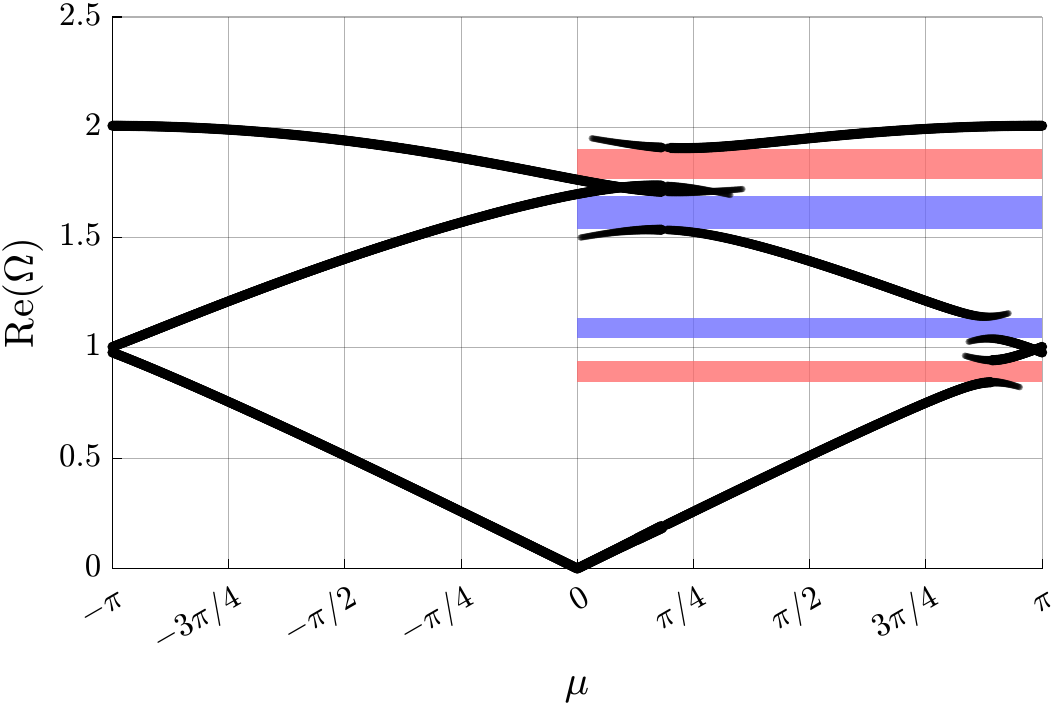}
        \caption{$\mathrm{Re}(\Omega)$ for $\alpha_m = 0.2$, $\Omega_m = 0.2$}
    \end{subfigure}
    \caption{Dispersion diagrams for SP system ($\alpha_m = 0.2$, $\Omega_m = 0$) and STM system ($\alpha_m = 0.2$, $\Omega_m = 0.2$)}  
    \label{fig: stm_dispersion_real_comparisson}
\end{figure*}

\begin{figure*}
    \centering
    \begin{subfigure}{0.44\textwidth}
        \centering
        \includegraphics[width=\textwidth]{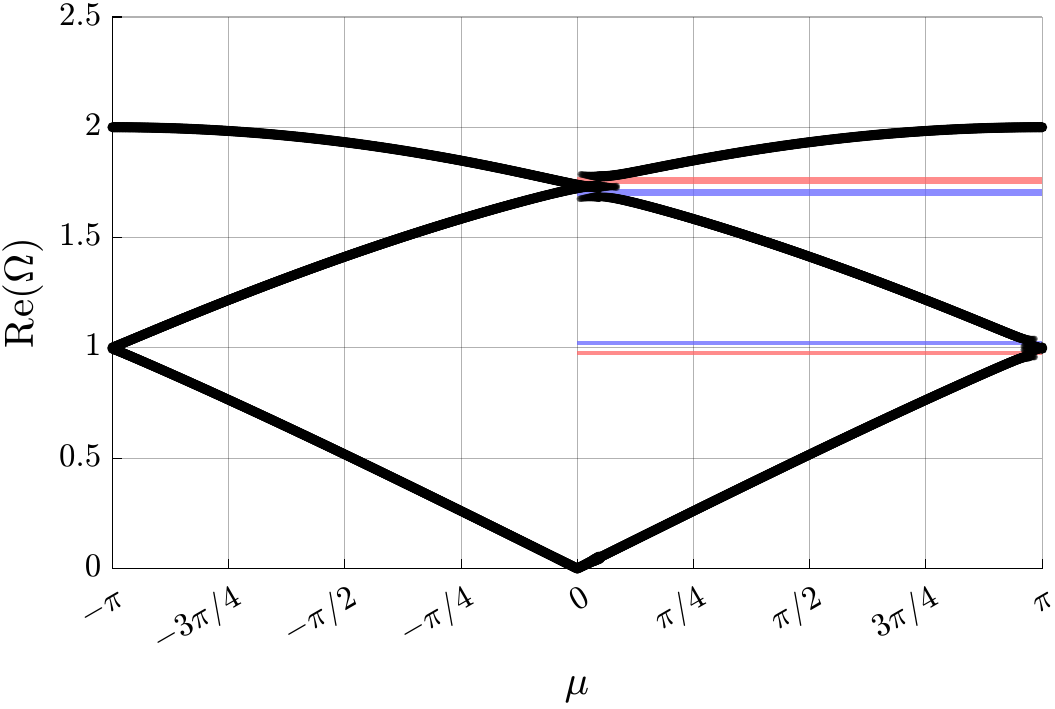}
        \caption{$\mathrm{Re}(\Omega)$ for $\alpha_m = 0.05$, $\Omega_m = 0.05$}
    \end{subfigure}\hfill
    \begin{subfigure}{0.44\textwidth}
        \centering
        \includegraphics[width=\textwidth]{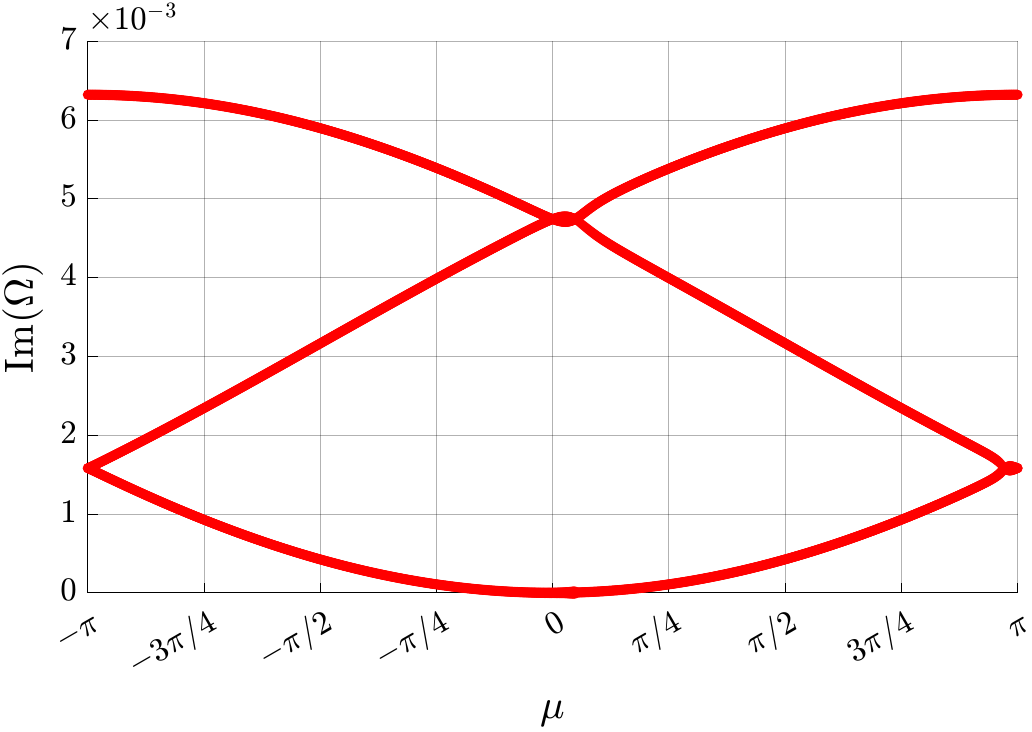}
        \caption{$\mathrm{Im}(\Omega)$ for $\alpha_m = 0.05$, $\Omega_m = 0.05$}
    \end{subfigure}
    \caption{STM dispersion diagrams for the case $(\alpha_m,\,\Omega_m)=(0.05,\,0.05)$, shown in terms of $\mathrm{Re}(\Omega)$ and $\mathrm{Im}(\Omega)$.}
    \label{fig: stm_dispersion_case_005_005}
\end{figure*}

\begin{figure*}
    \centering
    \begin{subfigure}{0.44\textwidth}
        \centering
        \includegraphics[width=\textwidth]{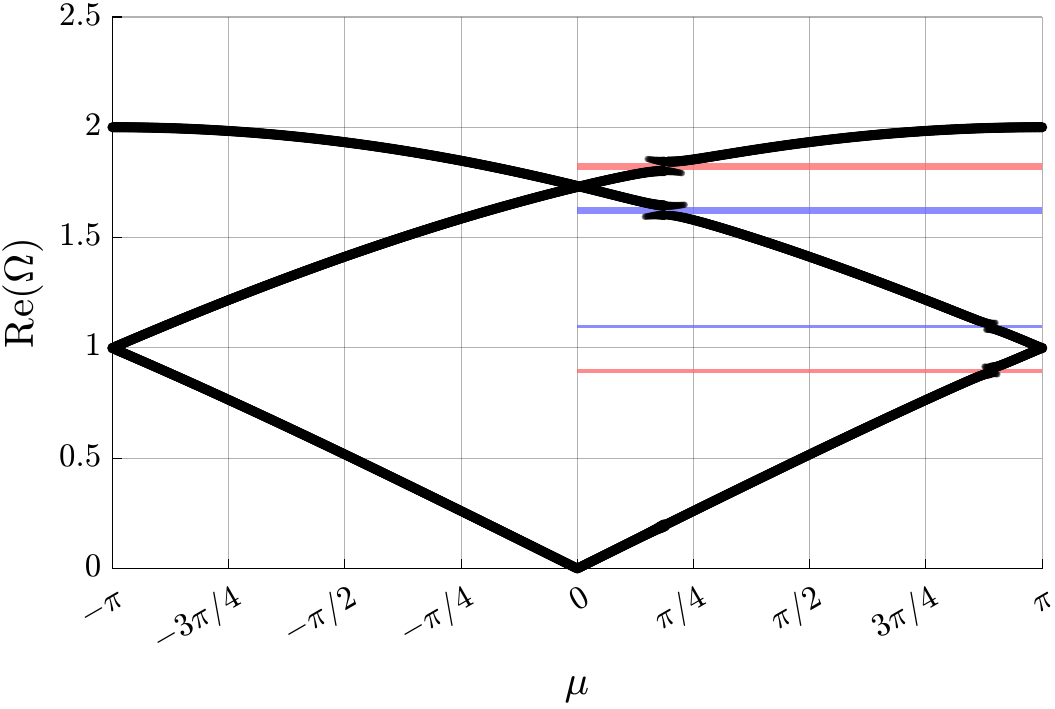}
        \caption{$\mathrm{Re}(\Omega)$ for $\alpha_m = 0.05$, $\Omega_m = 0.2$}
    \end{subfigure}\hfill
    \begin{subfigure}{0.44\textwidth}
        \centering
        \includegraphics[width=\textwidth]{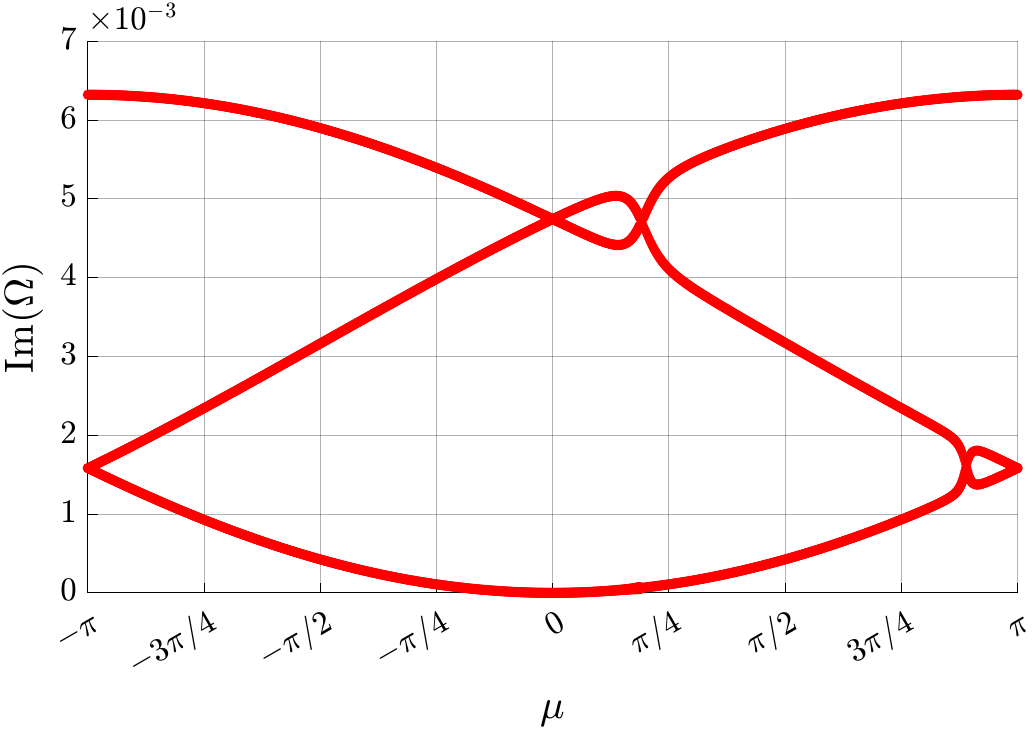}
        \caption{$\mathrm{Im}(\Omega)$ for $\alpha_m = 0.05$, $\Omega_m = 0.2$}
    \end{subfigure}
    \caption{STM dispersion diagrams for the case $(\alpha_m,\,\Omega_m)=(0.05,\,0.2)$, shown in terms of $\mathrm{Re}(\Omega)$ and $\mathrm{Im}(\Omega)$.}
    \label{fig: stm_dispersion_case_005_02}
\end{figure*}

\begin{figure*}
    \centering
    \begin{subfigure}{0.44\textwidth}
        \centering
        \includegraphics[width=\textwidth]{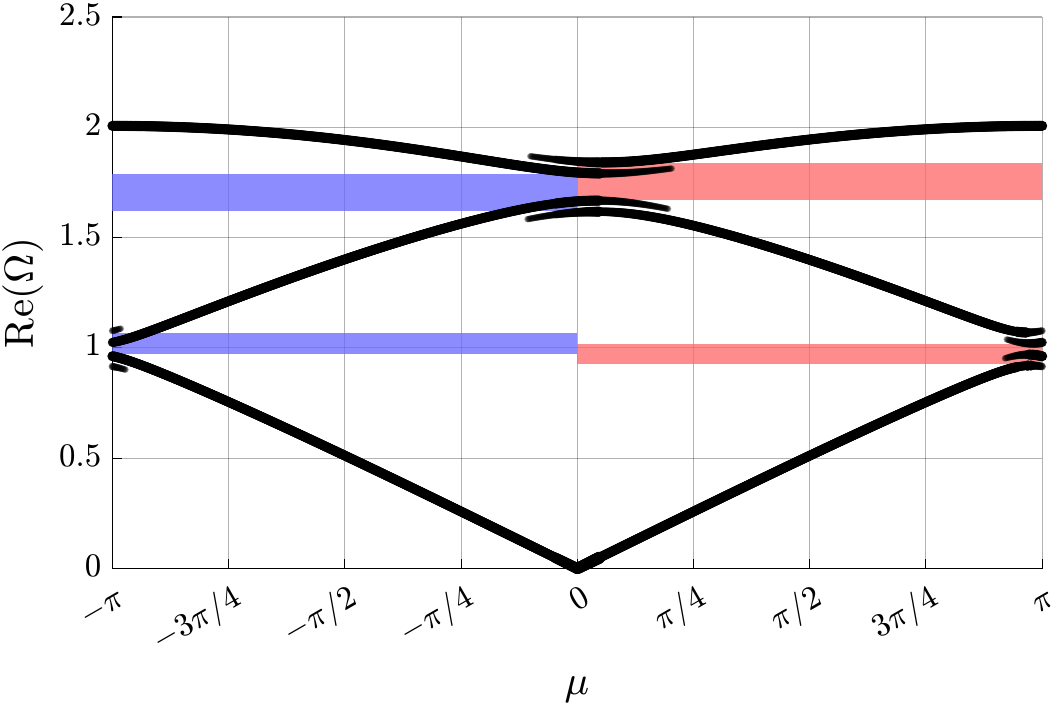}
        \caption{$\mathrm{Re}(\Omega)$ for $\alpha_m = 0.2$, $\Omega_m = 0.05$}
    \end{subfigure}\hfill
    \begin{subfigure}{0.44\textwidth}
        \centering
        \includegraphics[width=\textwidth]{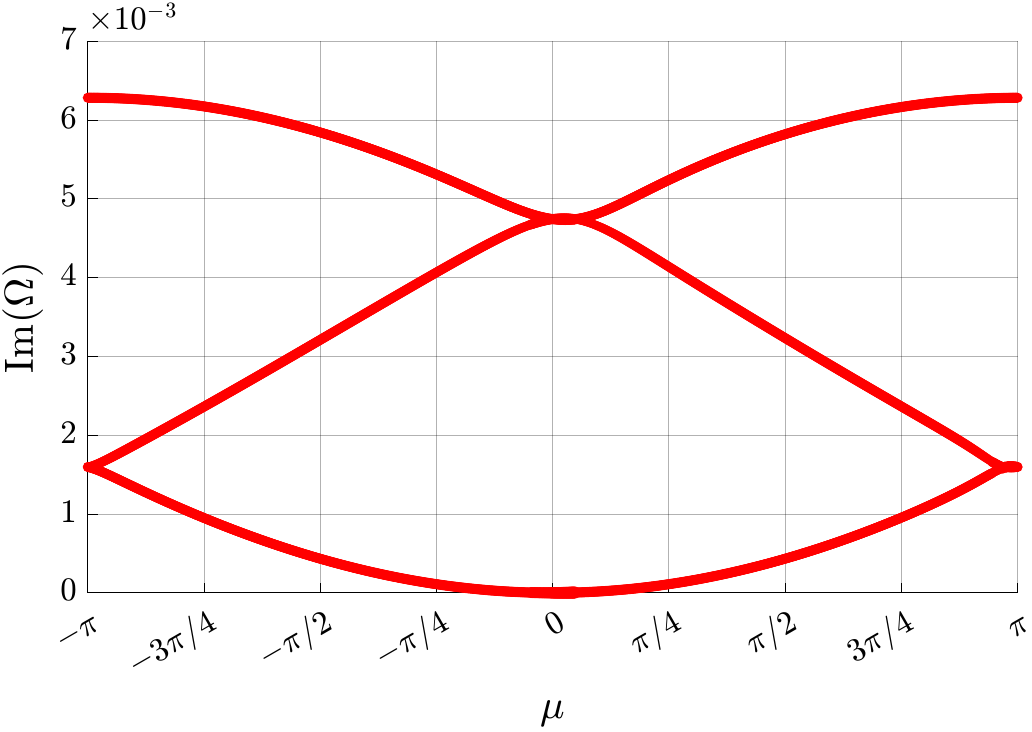}
        \caption{$\mathrm{Im}(\Omega)$ for $\alpha_m = 0.2$, $\Omega_m = 0.05$}
    \end{subfigure}
    \caption{STM dispersion diagrams for the case $(\alpha_m,\,\Omega_m)=(0.2,\,0.05)$, shown in terms of $\mathrm{Re}(\Omega)$ and $\mathrm{Im}(\Omega)$.}
    \label{fig: stm_dispersion_case_02_005}
\end{figure*}

\begin{figure*}
    \centering
    \begin{subfigure}{0.44\textwidth}
        \centering
        \includegraphics[width=\textwidth]{Figures/Dispersion/dispersion_stm_real_R3_P1_Om0.2_alpha0.2.pdf}
        \caption{$\mathrm{Re}(\Omega)$ for $\alpha_m = 0.2$, $\Omega_m = 0.2$}
    \end{subfigure}\hfill
    \begin{subfigure}{0.44\textwidth}
        \centering
        \includegraphics[width=\textwidth]{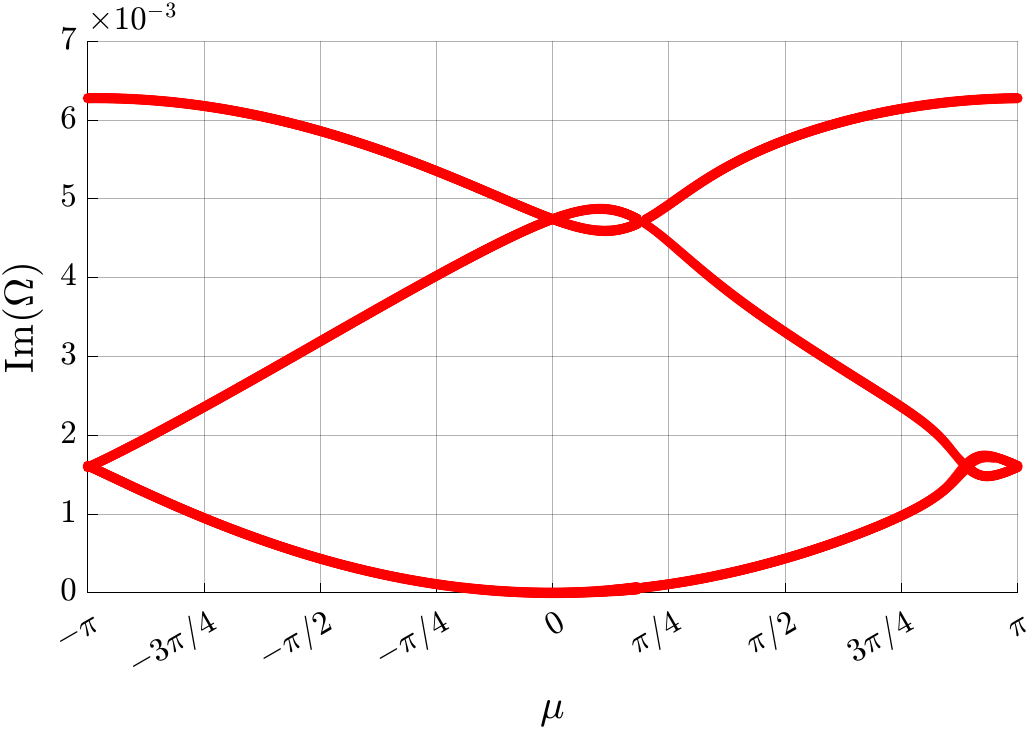}
        \caption{$\mathrm{Im}(\Omega)$ for $\alpha_m = 0.2$, $\Omega_m = 0.2$}
    \end{subfigure}
    \caption{STM dispersion diagrams for the case $(\alpha_m,\,\Omega_m)=(0.2,\,0.2)$, shown in terms of $\mathrm{Re}(\Omega)$ and $\mathrm{Im}(\Omega)$.}
    \label{fig: stm_dispersion_case_02_02}
\end{figure*}

Figure~\ref{fig: stm_dispersion_real_comparisson} compares real parts of the dispersion diagrams for the Spatially Modulated (SP) system ($\alpha_m = 0.2$, $\Omega_m = 0$) and the STM system ($\alpha_m = 0.2$, $\Omega_m = 0.2$). Figures~\ref{fig: stm_dispersion_case_005_005}--\ref{fig: stm_dispersion_case_02_02} shows $\mathrm{Re}(\Omega)$ and $\mathrm{Im}(\Omega)$ of dispersion diagram for a given modulation parameter pair. On those diagrams, the direction of band gaps is indicated by colored shaded areas, with blue and red corresponding to band gaps for negative and positive propagation directions, respectively. To analyze the STM dispersion diagrams, it is important to note that the modulation frequency $\Omega_m$ introduces asymmetry in the dispersion branches, yielding band gaps for $\mu \neq 0$ and $\mu \neq \pm \pi$. In this case, the slope $\xi = \partial \Omega/\partial \mu$ of the dispersion branches, which is related to the group velocity of the wave modes, can be used to identify propagation direction. For $\xi < 0$, the wave mode propagates in the negative direction, whereas for $\xi > 0$, it propagates in the positive direction. 


The dispersion diagram for the feedback system with $\gamma^v = -1$ reveals differences between the imaginary parts of the dispersion branches for $\mu > 0$ and $\mu < 0$ in the normalized frequency range $\Omega < 0.35$. In this interval, the imaginary part of the dispersion branches for $\mu < 0$ is significantly higher than that for $\mu > 0$, indicating that waves propagating in the negative direction are more attenuated than those propagating in the positive direction. This behavior leads to different effective propagation characteristics in opposite directions. As mentioned above, the projection of the feedback system dispersion diagram onto the complex plane forms a closed loop that encloses a nonzero area for sufficiently negative values of $\gamma^v$, indicating non-trivial non-Hermitian topology. The value of the velocity feedback gain $\gamma^v$ allows one to change the loop size and, therefore, the degree of non-trivial topology, a characteristic feature of feedback systems that is not observed in space-time modulated systems. A limitation of the system in the case of equal masses and springs is its single-band dispersion structure. Owing to this single-band behavior, our analysis of the non-reciprocal wave propagation characteristic is limited to $\Omega < 0.35$ for waves traveling in the negative direction. In contrast, it extends to $\Omega < 3.5$ for waves traveling in the positive direction. For this reason, the excitation frequencies $\Omega_f \in \{0.1,\, 0.3\}$ are selected to lie within the passband in both directions while preserving differences in the imaginary parts of the dispersion branches, thereby highlighting the non-reciprocal behavior.

For the STM system, the dispersion diagram reveals directional band gaps for both propagation directions. Considering $\alpha_m = 0.2$ and $\Omega_m = 0.2$, directional band gaps are observed for $\mu > 0$ within the normalized frequency ranges $\mathrm{Re}(\Omega) \in [0.84,\, 0.94] \cup [1.76,\, 1.90]$, while for $\mu < 0$ they appear in the ranges $\mathrm{Re}(\Omega) \in [1.04,\, 1.14] \cup [1.56,\, 1.70]$. Although some asymmetry is also visible in the imaginary part of the dispersion branches, its magnitude is small and does not produce differences as significant as those induced by the directional band gaps. For this reason, the dispersion diagrams are presented in both $\mathrm{Re}(\Omega)$ and $\mathrm{Im}(\Omega)$, with $\mathrm{Re}(\Omega)$ providing a clearer visualization of the directional band gaps. The excitation frequencies $\Omega_f \in \{1.6,\, 1.8\}$ are selected to highlight the non-reciprocal wave propagation characteristics of the system. An interesting feature of the space-time modulated system is that the modulation frequency $\Omega_m \neq 0$ causes the band gaps of the spatially modulated system to split into two separate band gaps, one for $\mu > 0$ and another for $\mu < 0$, with frequency intervals separated by $\Omega_m$. This behavior enables tuning of the directional band gap locations by adjusting the natural and modulation frequencies of the system, a unique feature of space-time modulated systems that is not observed in feedback systems.

\subsection{Time-domain response}
The time-domain response is analyzed to verify the presence of directional band gaps and non-reciprocal wave propagation characteristics predicted by the dispersion diagram, with the values $m = 1$~kg, $k_0 = 10$~N/m, and $b = 0.01$~N$\cdot$s/m. The finite structures comprise $N_c = 2N + 1 = 61$ unit cells ($N = 30$), with clamped (fixed) boundary conditions at both ends. To minimize spurious reflections at the boundaries, absorbing layers are introduced at both ends, in which the damping coefficient increases from its nominal value to $c_{\max} = 20$~N$\cdot$s/m following a cubic polynomial ramp. We consider a localized sine burst excitation at the central mass of the structure, with a carrier frequency $\Omega_f$, unit amplitude, and a Hanning window of width $N_{cy} = 50$ cycles, strategically chosen to highlight the effects observed in the dispersion diagrams. The total simulation time spans $2N_{cy}/\Omega_f$ dimensionless time units, corresponding to two times the burst duration. 

\begin{figure}[h!]
    \centering
    \begin{subfigure}{0.48\textwidth}
        \centering
        \includegraphics[width=\textwidth]{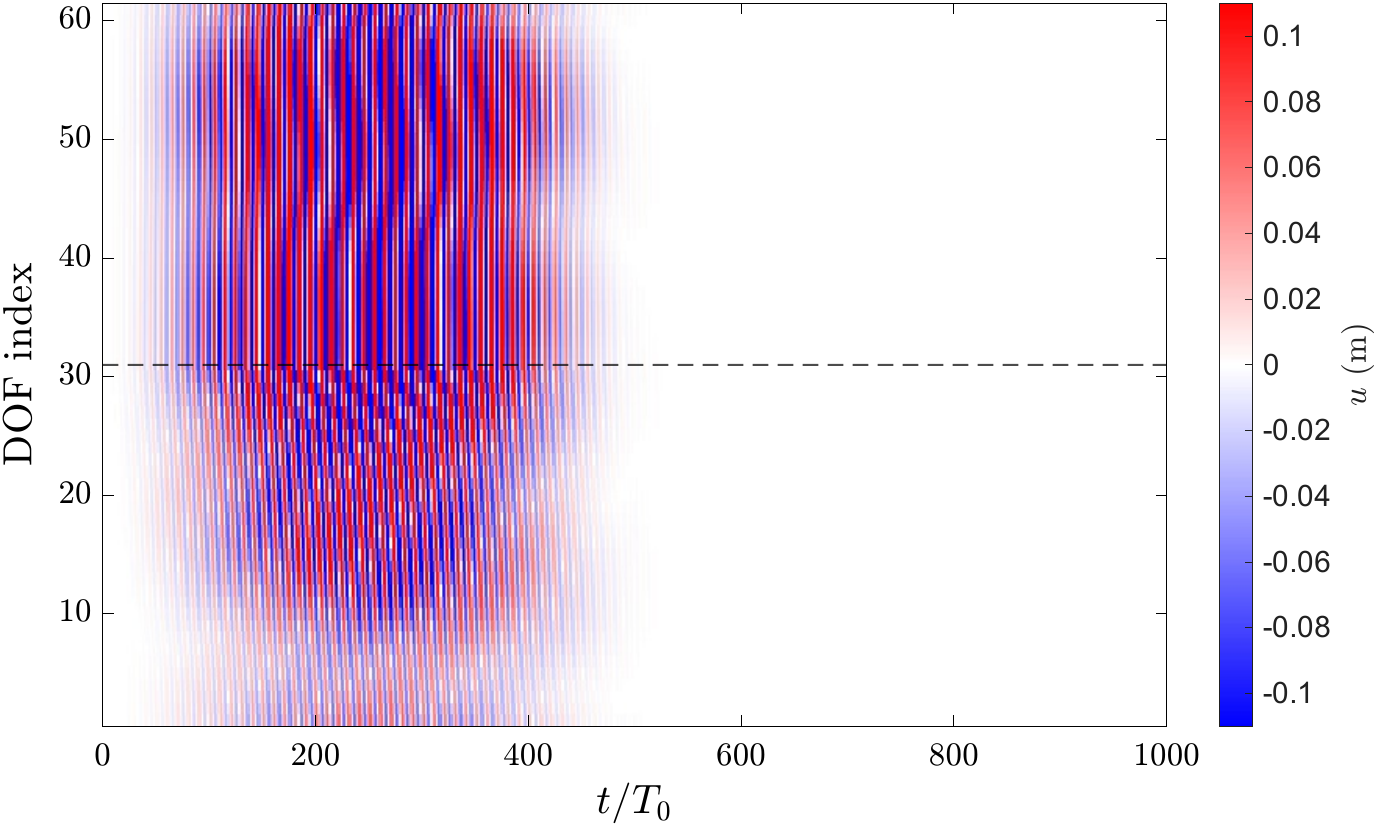}
        \caption{Time response of the feedback system for $\Omega_f = 0.1$}\label{fig: feedback_time_response_01a}
    \end{subfigure}
    \begin{subfigure}{0.48\textwidth}
        \centering
        \includegraphics[width=\textwidth]{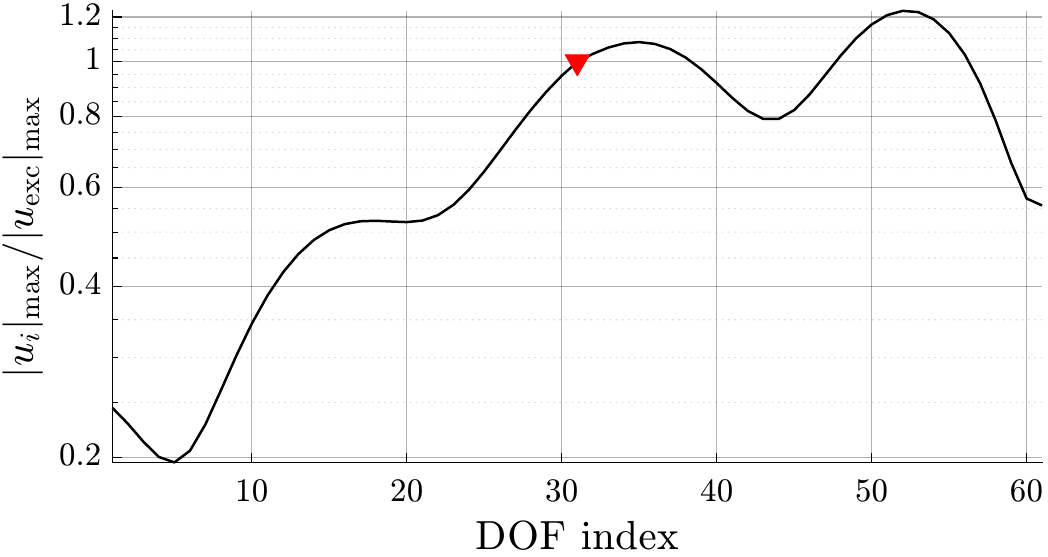}
        \caption{Maximum normalized displacement magnitude of the feedback system for $\Omega_f = 0.1$} \label{fig: feedback_time_response_01b}
    \end{subfigure}
    \begin{subfigure}{0.48\textwidth}
        \centering
        \includegraphics[width=\textwidth]{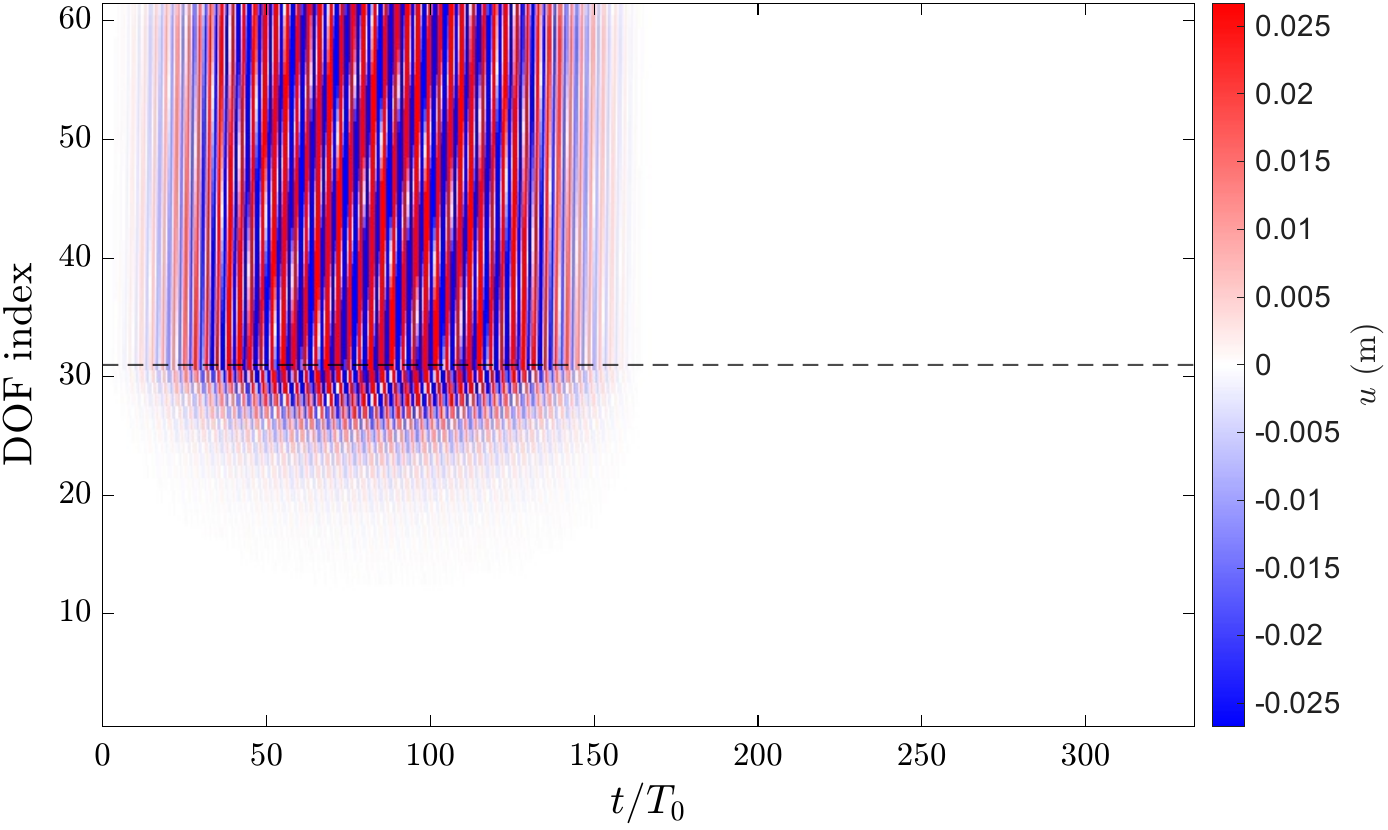}
        \caption{Time response of the feedback system for $\Omega_f = 0.3$}\label{fig: feedback_time_response_02a}
    \end{subfigure}
    \begin{subfigure}{0.48\textwidth}
        \centering
        \includegraphics[width=\textwidth]{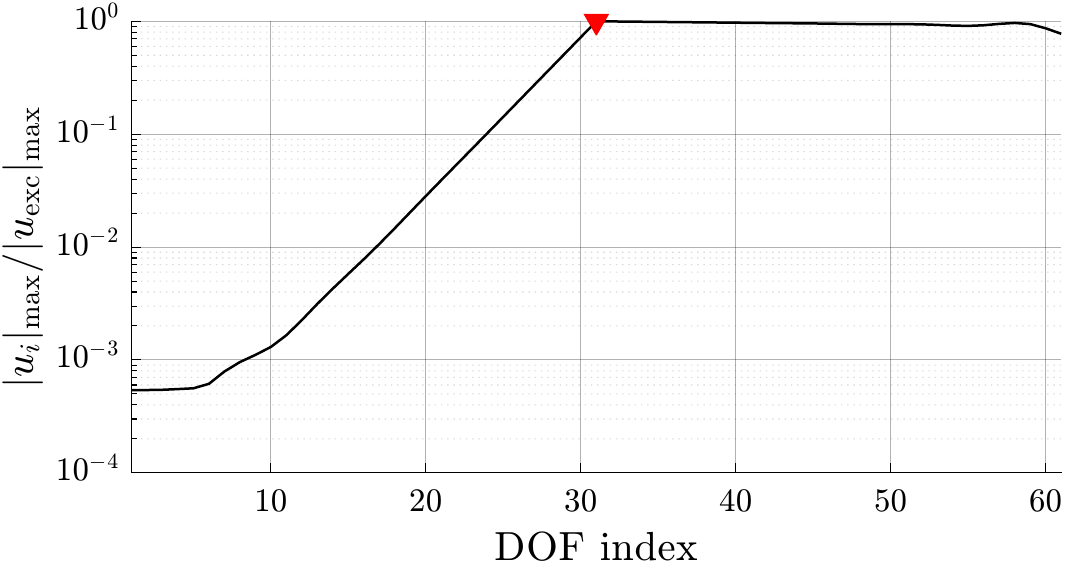}
        \caption{Maximum normalized displacement magnitude of the feedback system for $\Omega_f = 0.3$} \label{fig: feedback_time_response_02b}
    \end{subfigure}
    \caption{Time response of the feedback system for different excitation frequencies $\Omega_f$.}
\end{figure}

\begin{figure}[h!]
    \begin{subfigure}{0.48\textwidth}
        \centering
        \includegraphics[width=\textwidth]{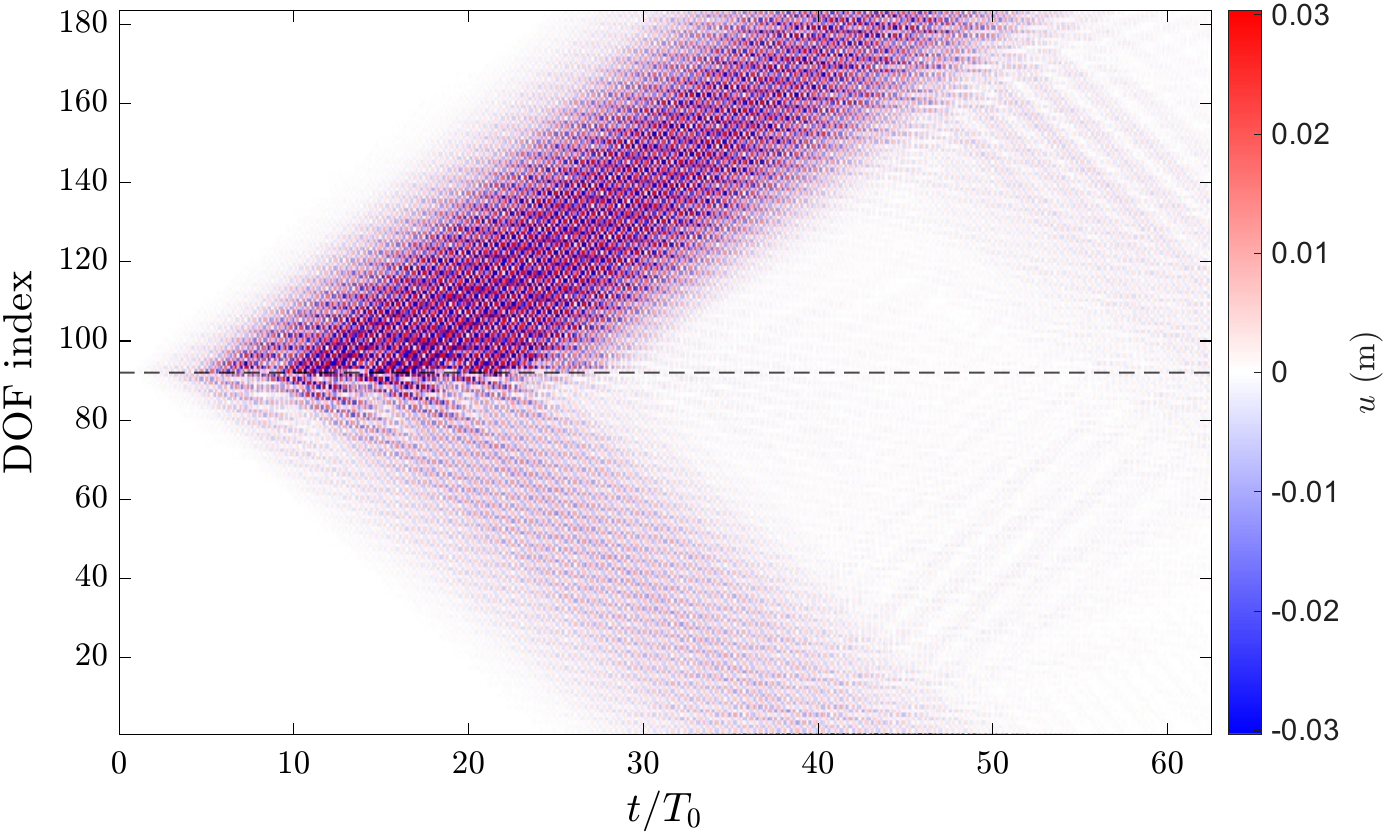}
        \caption{Time response of the STM system for $\Omega_f = 1.6$}\label{fig: stm_time_response_01a}
    \end{subfigure}
    \begin{subfigure}{0.48\textwidth}
        \centering
        \includegraphics[width=\textwidth]{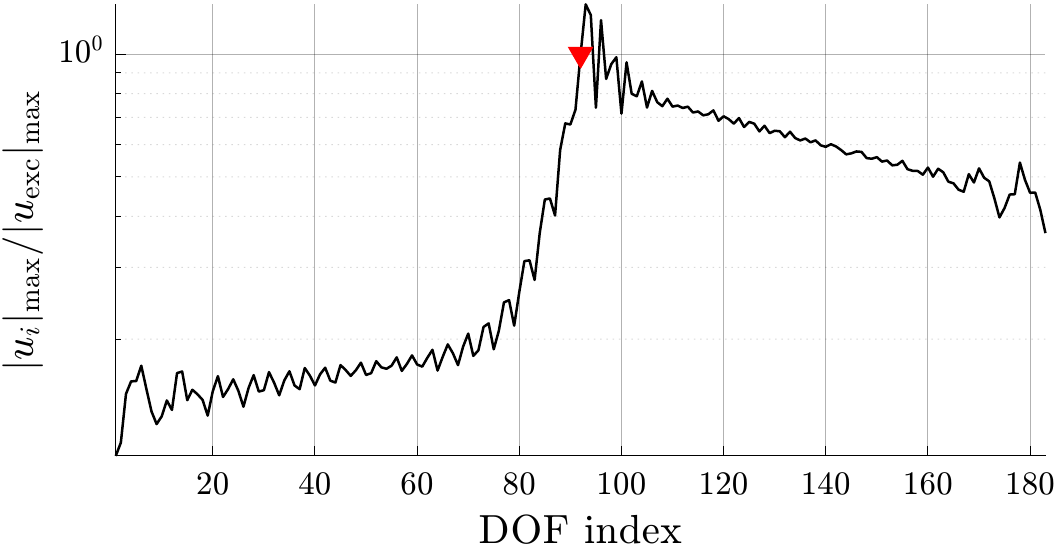}
        \caption{Maximum normalized displacement magnitude of the STM system for $\Omega_f = 1.6$} 
    \end{subfigure}\label{fig: stm_time_response_01b}
       \begin{subfigure}{0.48\textwidth}
        \centering
        \includegraphics[width=\textwidth]{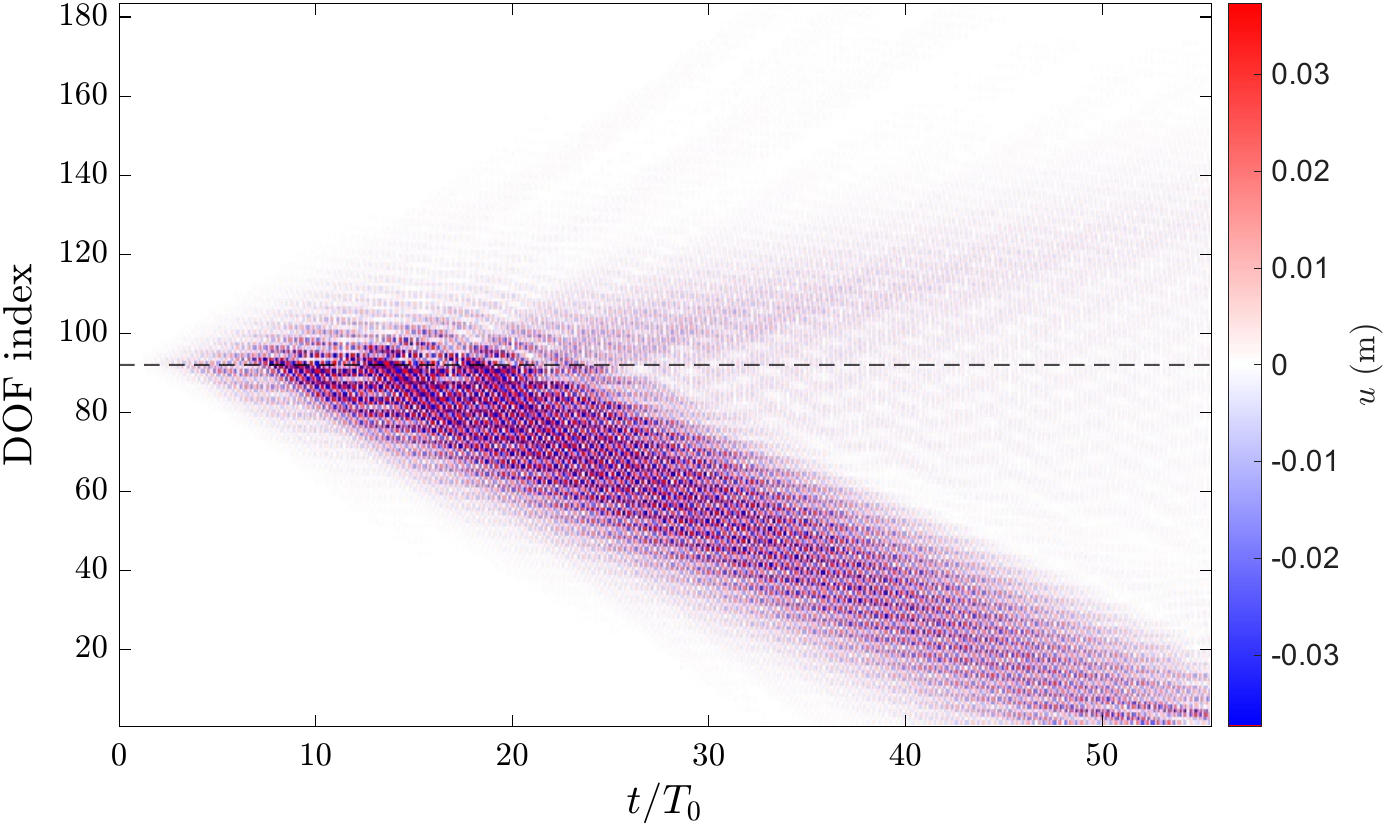}
        \caption{Time response of the STM system for $\Omega_f = 1.8$}\label{fig: stm_time_response_02a}
    \end{subfigure}
    \begin{subfigure}{0.48\textwidth}
        \centering
        \includegraphics[width=\textwidth]{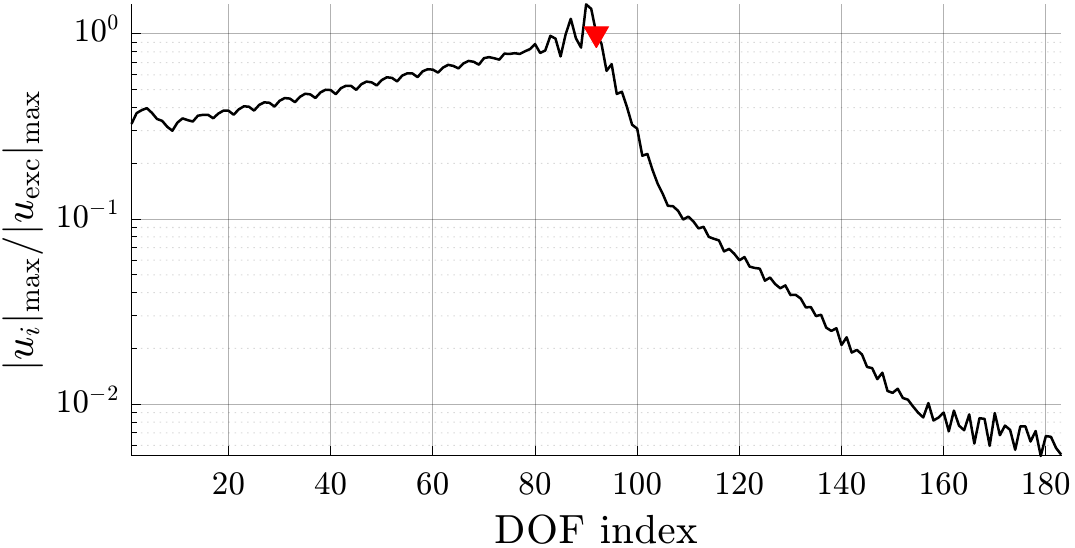}
        \caption{Maximum normalized displacement magnitude of the STM system for $\Omega_f = 1.8$} 
    \end{subfigure}\label{fig: stmm_time_response_02b}
    \caption{Time response of the STM system for different excitation frequencies $\Omega_f$.}
\end{figure}

The equations of motion are integrated over this time span with a time step $\Delta t = 10^{-3}\, \omega_0^{-1}$. For the feedback system, time integration is performed using an adaptive Runge--Kutta scheme (\textit{ode45} built-in function from MATLAB\textsuperscript{\tiny\textregistered}). 
The velocity feedback gain is set to $\gamma^v = -1$, with all other gains equal to zero ($\gamma^a = \gamma^d = 0$). For the space-time modulated system, the equations of motion are integrated using the Newmark average acceleration method ($\beta = 1/4$, $\gamma_N = 1/2$)~\cite{bathe1976numerical}. 
The modulation parameters are set to $\alpha_m = 0.2$ and $\Omega_m = 0.2$, with $R = 3$ degrees of freedom per unit cell. Figures~\ref{fig: feedback_time_response_01a} and \ref{fig: feedback_time_response_02a} show the time-domain response of the feedback system for excitation frequencies $\Omega_f = 0.1$ and $\Omega_f = 0.3$, respectively, while Figs.~\ref{fig: stm_time_response_01a} and \ref{fig: stm_time_response_02a} show the corresponding responses of the space-time modulated system for $\Omega_f = 1.6$ and $\Omega_f = 1.8$, respectively. To better compare the time-domain responses of the two systems, the maximum displacement magnitude of each mass is normalized to that of the central mass (the excited mass). The normalized maximum displacement magnitude is plotted as a function of mass index for all selected excitation frequencies.

For the feedback system, the response for $\Omega_f = 0.1$ shows significant attenuation of waves propagating in the negative direction compared with those propagating in the positive direction, consistent with the higher imaginary part of the dispersion branches for $\mu < 0$ in the corresponding frequency range. For $\Omega_f = 0.3$, the response shows an even more pronounced non-reciprocal behavior, with waves propagating in the negative direction being almost completely attenuated, while those propagating in the positive direction exhibit significant propagation. 

For the space-time modulated system, the time-domain responses for $\Omega_f = 1.6$ and $\Omega_f = 1.8$ show significant attenuation of waves propagating in the negative and positive directions, respectively, consistent with the presence of directional band gaps in the corresponding frequency ranges. The response for $\Omega_f = 1.6$ shows significant attenuation of waves propagating in the negative direction, while those propagating in the positive direction exhibit significant propagation. Conversely, the behavior is reversed for $\Omega_f = 1.8$.

\section{CONCLUSIONS}

This paper presents a comprehensive analysis of non-reciprocal wave propagation induced by feedback and space-time modulation (STM) in spatially periodic systems, revealing the critical role of feedback gains and modulation parameters in determining the stability of finite structures.  
The dispersion diagrams
display non-reciprocal wave propagation 
due to asymmetric branches in the systems with feedback, and directional band gaps in the STM systems, 
further validated by the time-domain simulations.
~The feedback approach offers simpler implementation, 
whereas 
obtaining the STM dispersion diagrams requires identifying the correct plane wave solutions. 
We observed that the directional band gaps exhibited by the STM systems cover narrow frequency intervals, 
whereas the feedback system exhibits non-reciprocal wave propagation for any frequency within topologically protected passbands.

The stability analysis of the STM systems has a higher computational cost due to the need to numerically compute the monodromy matrix and its eigenvalues
, whereas the analysis of the feedback systems only 
requires computing the eigenvalues of the dynamic matrix in the state-space representation. 
The feedback systems also 
remain stable for any number of unit cells, provided the feedback is velocity-based and has negative gain. In contrast, the space-time modulated system can become unstable for certain combinations of modulation parameters, and its stability region can shrink as the number of unit cells and/or the modulation amplitude and modulation frequency increase.

\section{ACKNOWLEDGMENTS}

The authors acknowledge the support of FINEP (Financiadora de Estudos e Projetos) through the project METABIO3D (Grant No. 1550/22) within the Rota2030 program, 
CNPq (Grant No. 305293/2021-4) and FAPESP (Grant No. 2018/15894-0). Danilo Braghini is thankful to the support of the National Science Foundation under grant No. 2337751.


\bibliographystyle{apalike}
\bibliography{bibfile}

\section{RESPONSIBILITY NOTICE}

The authors are solely responsible for the printed material included in this paper.

\end{document}